\documentclass{article}

\begin{document}

\centerline{The necessary and sufficient condition for perfect
teleportation and superdense coding}

\centerline{and all the suitable states for teleportation and superdense
coding}

\centerline{Dafa Li}
\centerline{Department of Mathematical Sciences,
Tsinghua University, Beijing 100084, CHINA}
\centerline{email:
lidafa@tsinghua.edu.cn}

Abstract.

It is known that two local unitaries (LU) equivalent states possess the same
amount of entanglement and can be used to perform the same tasks in quantum
information theory (QIT). For a protocol for a task in QIT, we call a
protocol LU invariant if two LU-equivalent states are either both suitable
for the protocol or neither is. So far, no one has discussed whether a
protocol for a task in QIT is LU invariant. In [Phys. Rev. A, 74, 062320
(2006)], Agrawal and Pati proposed the perfect teleportation protocol (PTP)
and the protocol for superdense coding to transmit 2-bit classical
information by sending one qubit (PSDC-2) and 3-bit classical information by
sending two qubits (PSDC-3). In this paper, we show that PTP and PSDC-2 are
LU invariant. That is, two LU equivalent states are suitable for PTP and
PSDC-2 or neither of them is. We show that PSDC-3 is not LU invariant. We
also indicate that the teleportation proposed in \cite{Nielsen} is not LU
invariant. We give a necessary and sufficient condition for a state to be
suitable for PTP, PSDC-2, and PSDC-3, respectively. Via the LU invariance of
PTP and PSDC-2, we prove that a state is suitable for PTP and PSDC-2 if and
only if it has 1 ebit of shared entanglement, respectively and find all
genuine entangled states and separable states which are suitable for PTP and
PSDC-2, respectively. So far, no one has indicated that PTP and PSDC-2 do
not require genuine entanglement. Agrawal and Pati suggested to study if
there are subclasses of W SLOCC class which are suitable for PSDC-3. So far,
it still remains an unsolved question. We show that any state of the SLOCC\
class W is not suitable for PSDC-3.


\section{Introduction}

SLOCC (LU) equivalence of two pure states $|\psi \rangle $ and $|\psi
^{\prime }\rangle $ of $n$ qubits is defined as $|\psi ^{\prime }\rangle
=A_{1}\otimes \cdots \otimes A_{n}|\psi \rangle $, where $A_{i}$ are local
invertible operators (local unitaries) \cite{Bennett, Dur, Verstraete,
Kraus-prl, Kraus-pra}. Numerous efforts have contributed to entanglement
classifications LU, LOCC, and SLOCC \cite{Bennett, Dur, Verstraete,
Kraus-prl, Kraus-pra, Miyake, Luque, LDFQIC09, Leifer, Osterloh06, Levay,
LDF07a, Bastin, LDFPRL12, LDFPRA12, Walter}. For example, D\"{u}r et al.
partitioned three qubits into six SLOCC classes GHZ, W, A-BC, B-AC, C-AB,
and A-B-C \cite{Dur}. Any pure state of three qubits can be transformed into
the Schmidt decomposition (SD) under LU \cite{Acin00}. In \cite{Dli-qip-18,
Dli-jpa-20, Dli-qic-23}, we classified the Schmidt decompositions of three
qubits.

The original quantum teleportation is to transmit an unknown state perfectly
using Bell states and 2-bit classical communication between the sender and
the receiver \cite{Bennett}. In \cite{Pankaj},\ Agrawal and Pati proposed
PTP, PSDC-2, and PSDC-3, respectively. They indicated that W state cannot be
suitable for PTP, PSDC-2, or PSDC-3. They stated that the subclass $%
|W_{n}\rangle _{123}$\ of W SLOCC\ class is suitable for PTP and PSDC-2. Li
and Qiu also studied the states of W-class as shared resources for perfect
teleportation and superdense coding and found a subclass of W-class\ being
suitable for their perfect teleportation and superdense coding \cite{LLI}.
Previous articles have made many efforts to explore if there are other
subclasses of W SLOCC\ class which are suitable for PTP and PSDC-2. W-like
states are more general quantum entanglement states used in many quantum
communication schemes, distillation, teleportation and superdense coding
\cite{Lo, Ding, Pankaj, LLI, Singh}.

The maximally entangled states were used as resource states for
teleportation and superdense coding \cite{LM, VN, SB, XS} and non-maximally
entangled states were used as resource states for probabilistic
teleportation and probabilistic superdense coding \cite{WL, GG, SM, AP-05,
Singh}. Singh et al. indicated that the use of non-maximally entangled
resources leads to probabilistic protocols and the fidelity of information
transfer is always less than unity \cite{Singh}.

In this paper, we propose a necessary and sufficient condition for\ PTP,
PSDC-2, and PSDC-3, respectively. We show that PTP and PSDC-2 are LU
invariant and indicate that PSDC-3 and the teleportation in \cite{Nielsen}\
are not LU invariant. Via the LU invariance, we prove that a state of three
qubits is suitable for PTP and PSDC-2 if and only if it has 1 ebit of shared
entanglement. We also indicate that PTP and PSDC-2 do not require genuine
entanglement.

Notation. Let $|\psi \rangle _{123}$ be a state of three qubits 1, 2, and 3
and $\rho _{123}$ be the density matrix of $|\psi \rangle _{123}$. We write $%
|\psi \rangle _{123}$ as $|\psi \rangle $ for simplicity sometimes. Let $%
\rho _{1}=tr_{23}\rho _{123}$, $\rho _{2}=tr_{13}\rho _{123}$, and $\rho
_{3}=tr_{12}\rho _{123}$.

von Neumann entanglement entropy is defined as
\begin{equation}
S(\rho )=-\sum \eta _{i}\ln \eta _{i},  \label{von-1}
\end{equation}%
where $\eta _{i}\geq 0$ are the eigenvalues of $\rho $, and $\sum_{i}\eta
_{i}=1$. Note that $0\ln 0=0$. GHZ state is $\frac{1}{\sqrt{2}}(|000\rangle
+|111\rangle )$ and $S(\rho _{i}(|GHZ\rangle )=\ln 2$, $i=1,2,3$. W state is
$\frac{1}{\sqrt{3}}(|001\rangle +|010\rangle +|100\rangle )$ and $S(\rho
_{i}(|W\rangle )=\frac{3\ln 3-2\ln 2}{3}$, $i=1,2,3$.

\section{For PTP}

Agrawal and Pati introduced how to use the three-qubit GHZ state as the
resource state for PTP. They also indicated that the following three-qubit
state $|W_{n}\rangle _{123}$\ are suitable for PTP \cite{Pankaj}.

\begin{equation}
|W_{n}\rangle _{123}=\frac{1}{\sqrt{2}\sqrt{n+1}}(|100\rangle +\sqrt{n}%
e^{i\gamma }|010\rangle +\sqrt{n+1}e^{i\delta }|001\rangle ).  \label{W-n}
\end{equation}

In this section, we give a necessary and sufficient condition for a state of
$n$ qubits to be suitable for PTP and find all the suitable states.

\subsection{ For a state to be suitable for PTP}

In this section, we want to find all the states which are suitable for PTP.
Let $|\psi \rangle _{123}$ be any state of three qubits. Assume that $|\psi
\rangle _{123}$ is suitable for PTP. Let Alice and Bob share the state $%
|\psi \rangle _{123}$. Alice has qubits 1 and 2 while Bob has qubit 3. Alice
has another qubit \textquotedblleft $a$\textquotedblright\ in the unknown
state $|\varphi \rangle _{a}=(\alpha |0\rangle _{a}+\beta |1\rangle _{a})$.
Alice wants to send the unknown state $|\varphi \rangle _{a}$ to Bob. She
needs to perform a von Neumann joint measurement on the three qubits `a12'\
by using a set of orthogonal states\ and then conveys her result to Bob by
sending 2-bit classical message. Then, Bob can convert his state to the
original unknown state\ by applying appropriate unitary transformations.
Thus, this completes the teleportation protocol using the state $|\psi
\rangle _{123}$.

\subsection{PTP is LU invariant}

It was known that two LU equivalent states possess the same amount of
entanglement and can be used to do the same tasks in QIT \cite{Dur,
Verstraete, Kraus-prl, Kraus-pra}. We call a protocol for a task in QIT LU
invariant if a state can be used for the protocol for the task then any
state being LU equivalent to the state can also be used for the same
protocol. If a protocol for a task in QIT is not LU invariant, then there
are two LU equivalent states such that one of which can be used for the
protocol while the other one cannot.

In this subsection, we explore if PTP is LU invariant. To this end, we need
the schmidt decomposition (SD) of three qubits.

For any three-qubit state $|\psi \rangle _{123}$, there are local unitary
operators $U_{i}$, $i=1,2,3$, such that $|\psi \rangle _{123}$ can be
transformed into SD
\begin{eqnarray}
&&|\psi ^{SD}\rangle _{123}=U_{1}\otimes U_{2}\otimes U_{3}|\psi \rangle
_{123}  \nonumber \\
&=&(\lambda _{0}|000\rangle +\lambda _{1}e^{i\phi }|100\rangle +\lambda
_{2}|101\rangle +\lambda _{3}|110\rangle +\lambda _{4}|111\rangle )_{123},
\label{sd}
\end{eqnarray}%
where $\lambda _{i}\geq 0$, $i=0$ to $4$, $\sum_{i=0}^{4}\lambda _{i}^{2}=1$%
, and $\phi $ is called a phase \cite{Acin00}. In this paper, $0\leq \phi
<2\pi $. Note that the phase $\phi $ is limited to $[0,\pi ]$ in \cite%
{Acin00}.

It is clear that a state is LU equivalent to its SD. The following theorem
confirms that PTP is LU invariant.

Theorem 1. $|\psi \rangle $ is suitable for PTP if and only if its SD $|\psi
^{SD}\rangle $ is suitable for PTP.

The proof is put in Appendix A. Via the LU invariance for PTP, we will find
all the states which are suitable for PTP.

\subsection{A necessary and sufficient condition for PTP}

In this section, we want to find all the states which are suitable for PTP.
Let $|\psi \rangle _{123}=\sum_{i=0}^{7}c_{i}|i\rangle _{123}$, where $%
\sum_{i=0}^{7}|c_{i}|^{2}=1$, be any pure state of three qubits and $%
|\varphi \rangle _{a}=\alpha |0\rangle _{a}+\beta |1\rangle _{a}$. We can
also write $|\psi \rangle _{123}=\sum_{ijk=0,1}c_{ijk}|ijk\rangle _{123}$.
Let
\begin{eqnarray}
|\Xi \rangle _{12} &=&(c_{0}|00\rangle +c_{2}|01\rangle +c_{4}|10\rangle
+c_{6}|11\rangle )_{12},  \nonumber \\
|\Pi \rangle _{12} &=&(c_{1}|00\rangle +c_{3}|01\rangle +c_{5}|10\rangle
+c_{7}|11\rangle )_{12}.
\end{eqnarray}%
Via the above $|\Xi \rangle _{12}$ and $|\Pi \rangle _{12}$, $|\psi \rangle
_{123}$ can be written as
\begin{equation}
|\psi \rangle _{123}=|\Xi \rangle _{12}|0\rangle _{3}+|\Pi \rangle
_{12}|1\rangle _{3}.
\end{equation}%
Then, obtain

\begin{eqnarray}
|\varphi \rangle _{a}|\psi \rangle _{123} &=&(\alpha |0\rangle _{a}+\beta
|1\rangle _{a})(|\Xi \rangle _{12}|0\rangle _{3}+|\Pi \rangle _{12}|1\rangle
_{3})  \nonumber \\
&=&\alpha |0\rangle _{a}|\Xi \rangle _{12}|0\rangle _{3}+\alpha |0\rangle
_{a}|\Pi \rangle _{12}|1\rangle _{3}  \nonumber \\
&&+\beta |1\rangle _{a}|\Xi \rangle _{12}|0\rangle _{3}+\beta |1\rangle
_{a}|\Pi \rangle _{12}|1\rangle _{3}  \label{tel-2}
\end{eqnarray}

Let
\begin{eqnarray}
|\chi ^{\pm }\rangle _{a12} &=&|0\rangle _{a}|\Xi \rangle _{12}\pm |1\rangle
_{a}|\Pi \rangle _{12}, \\
|\varsigma ^{\pm }\rangle _{a12} &=&|1\rangle _{a}|\Xi \rangle _{12}\pm
|0\rangle _{a}|\Pi \rangle _{12}.
\end{eqnarray}%
Then,
\begin{eqnarray}
|\chi ^{+}\rangle _{a12}+|\chi ^{-}\rangle _{a12} &=&2|0\rangle _{a}|\Xi
\rangle _{12},  \label{orth-1} \\
|\chi ^{+}\rangle _{a12}-|\chi ^{-}\rangle _{a12} &=&2|1\rangle _{a}|\Pi
\rangle _{12},  \label{orth-2} \\
|\varsigma ^{+}\rangle _{a12}+|\varsigma ^{-}\rangle _{a12} &=&2|1\rangle
_{a}|\Xi \rangle _{12},  \label{orth-3} \\
|\varsigma ^{+}\rangle _{a12}-|\varsigma ^{-}\rangle _{a12} &=&2|0\rangle
_{a}|\Pi \rangle _{12}.  \label{orth-4}
\end{eqnarray}%
Then, from Eqs. (\ref{tel-2}, \ref{orth-1}, \ref{orth-2}, \ref{orth-3}, \ref%
{orth-4}), obtain
\begin{eqnarray*}
&&|\varphi \rangle _{a}|\psi \rangle _{123} \\
&=&\frac{1}{2}[\alpha (|\chi ^{+}\rangle _{a12}+|\chi ^{-}\rangle
_{a12})|0\rangle _{3}+\alpha (|\varsigma ^{+}\rangle _{a12}-|\varsigma
^{-}\rangle _{a12})|1\rangle _{3} \\
&&+\beta (|\varsigma ^{+}\rangle _{a12}+|\varsigma ^{-}\rangle
_{a12})|0\rangle _{3}+\beta (|\chi ^{+}\rangle _{a12}-|\chi ^{-}\rangle
_{a12})|1\rangle _{3}] \\
&=&\frac{1}{2}[|\chi ^{+}\rangle _{a12}(\alpha |0\rangle _{3}+\beta
|1\rangle _{3})+|\chi ^{-}\rangle _{a12}(\alpha |0\rangle _{3}-\beta
|1\rangle _{3}) \\
&&+|\varsigma ^{+}\rangle _{a12}(\beta |0\rangle _{3}+\alpha |1\rangle
_{3})+|\varsigma ^{-}\rangle _{a12}(\beta |0\rangle _{3}-\alpha |1\rangle
_{3})].
\end{eqnarray*}

Note that $\sigma _{3}(\alpha |0\rangle _{3}-\beta |1\rangle _{3})=\sigma
_{1}(\beta |0\rangle _{3}+\alpha |1\rangle _{3})=$ $\sigma _{1}\sigma
_{3}(\beta |0\rangle _{3}-\alpha |1\rangle _{3})=\alpha |0\rangle _{3}+\beta
|1\rangle _{3}$, where $\sigma _{1}\sigma _{3}=-i\sigma _{2}$. Note that
\textquotedblleft $-$\textquotedblright\ in Eq. (13) in \cite{Pankaj}\
should be deleted.

For perfect teleportation, it needs to guarantee that $|\chi ^{+}\rangle
_{a12}$, $|\chi ^{-}\rangle _{a12}$, $|\varsigma ^{+}\rangle _{a12}$, and $%
|\varsigma ^{-}\rangle _{a12}$\ are a set of orthogonal states. To this end,
the dot product of any two of these states must vanish. After receiving
2-bit classical information from Alice, Bob can convert his state to the
original unknown state\ by applying the unitary transformations in \{$I,$ $%
\sigma _{1},\sigma _{3},\sigma _{1}\sigma _{3}$\}.

Let us compute dot products of the 6 pairs of states now. A calculation
yields
\begin{equation}
\langle \chi ^{+}|\chi ^{-}\rangle =\langle \varsigma ^{+}|\varsigma
^{-}\rangle =\sum_{i=0}^{3}|c_{2i}|^{2}-\sum_{i=0}^{3}|c_{2i+1}|^{2}.
\label{coe-1}
\end{equation}%
From $\sum_{i=0}^{7}|c_{i}|^{2}=1$ and Eq. (\ref{coe-1}), one can see that $%
\langle \chi ^{+}|\chi ^{-}\rangle =\langle \varsigma ^{+}|\varsigma
^{-}\rangle =0$ if and only if
\begin{equation}
\sum_{i=0}^{3}|c_{2i}|^{2}=\sum_{i=0}^{3}|c_{2i+1}|^{2}=\frac{1}{2}.
\label{cod-1}
\end{equation}

A calculation yields that $\langle \chi ^{-}|\varsigma ^{-}\rangle =-\langle
\chi ^{+}|\varsigma ^{+}\rangle $ and $\langle \chi ^{-}|\varsigma
^{+}\rangle =-\langle \chi ^{+}|\varsigma ^{-}\rangle $ and
\begin{eqnarray}
\langle \chi ^{+}|\varsigma ^{+}\rangle &=&\sum_{i=0}^{3}c_{2i}^{\ast
}c_{2i+1}+\sum_{i=0}^{3}c_{2i}c_{2i+1}^{\ast },  \label{cod-2} \\
\langle \chi ^{+}|\varsigma ^{-}\rangle &=&-\sum_{i=0}^{3}c_{2i}^{\ast
}c_{2i+1}+\sum_{i=0}^{3}c_{2i}c_{2i+1}^{\ast },  \label{cod-3}
\end{eqnarray}%
where $c_{i}^{\ast }$ is the complex conjugate of $c_{i}$.

From Eqs. (\ref{cod-2}, \ref{cod-3}), clearly $\langle \chi ^{+}|\varsigma
^{+}\rangle =\langle \chi ^{+}|\varsigma ^{-}\rangle =0$ if and only if
\begin{equation}
\sum_{i=0}^{3}c_{2i}c_{2i+1}^{\ast }=0  \label{cod-4}
\end{equation}

Thus, we can conclude the following.

Theorem 2. Let $|\psi \rangle =\sum_{i=0}^{7}c_{i}|i\rangle $ be any state
of three qubits. Then, $|\psi \rangle $ is suitable for PTP if and only if $%
|\psi \rangle $ satisfies the two conditions in Eqs. (\ref{cod-1}, \ref%
{cod-4}).

Ex. 1. It is easy to check that $|W_{n}\rangle _{123}$ in Eq. (\ref{W-n})
satisfies the two conditions in Eqs. (\ref{cod-1}, \ref{cod-4}). So, it
verifies that $|W_{n}\rangle _{123}$ is suitable for PTP.

Ex. 2. Let $|G\rangle =\frac{1}{2}(|000\rangle +|010\rangle +|101\rangle
-|111\rangle )$. It is easy to check that the state $\frac{1}{\sqrt{2}}%
(|101\rangle +|110\rangle )$ (separable), GHZ state, $|G\rangle $ and the
following two graph states\ satisfy the two conditions in Eqs. (\ref{cod-1}, %
\ref{cod-4}).\
\begin{eqnarray}
&&\frac{1}{2\sqrt{2}}(|000\rangle +|001\rangle +|010\rangle -|011\rangle
+|100\rangle +|101\rangle -|110\rangle +|111\rangle ),  \label{graph-1} \\
&&\frac{1}{2\sqrt{2}}(|000\rangle +|001\rangle +|010\rangle -|011\rangle
+|100\rangle -|101\rangle -|110\rangle -|111\rangle ).  \label{graph-2}
\end{eqnarray}

Ex. 3. W state and the general GHZ\ state $\alpha |000\rangle +\beta
|111\rangle $, where $|\alpha |\neq |\beta |$ and $|\alpha |^{2}+|\beta
|^{2}=1$, do not satisfy the condition in Eq. (\ref{cod-1}).

\subsection{PTP with SD}

For any three-qubit state $|\psi \rangle _{123}$, there are local unitary
operators $U_{i}$, $i=1,2,3$, such that $|\psi \rangle _{123}$ can be
transformed into SD in Eq. (\ref{sd}). In \cite{Dli-qip-18, Dli-jpa-20,
Dli-qic-23}, we classified the Schmidt decompositions of three qubits. From
Table 2 of \cite{Dli-qip-18}, we have the following SDs for SLOCC classes
GHZ, W, A-BC, B-AC, C-AB, respectively. In this subsection, we want to find
all SDs which are suitable for PTP. Then, via the LU invariance of PTP, we
get all the states which are suitable for PTP.

\subsubsection{Classification of SDs}

SD of the state belonging to GHZ SLOCC\ class is of the following form

\begin{eqnarray}
&&|\psi ^{SD}\rangle  \nonumber \\
&=&(\lambda _{0}|000\rangle +\lambda _{1}e^{i\phi }|100\rangle +\lambda
_{2}|101\rangle +\lambda _{3}|110\rangle )+\lambda _{4}|111\rangle ),
\label{ghz-sd}
\end{eqnarray}%
where $\lambda _{0}\lambda _{4}\neq 0$.

SD of the state belonging to W SLOCC\ class is of the following form.
\begin{equation}
|W^{SD}\rangle =\lambda _{0}|000\rangle +\lambda _{1}e^{i\phi }|100\rangle
+\lambda _{2}|101\rangle +\lambda _{3}|110\rangle ),  \label{w-sd}
\end{equation}%
where $\lambda _{0}\lambda _{2}\lambda _{3}\neq 0$.

SD of the state belonging to A-BC SLOCC\ class is of the following form.

\begin{equation}
|\psi ^{SD}\rangle =\lambda _{1}e^{i\phi }|100\rangle +\lambda
_{2}|101\rangle +\lambda _{3}|110\rangle )+\lambda _{4}|111\rangle ,
\label{a-bc-sd}
\end{equation}%
where $\lambda _{1}\lambda _{4}e^{i\phi }-\lambda _{2}\lambda _{3}\neq 0$.

SD of the state belonging to B-AC SLOCC\ class is of the following form

\begin{equation}
|\psi ^{SD}\rangle =\lambda _{0}|000\rangle +\lambda _{1}e^{i\varphi
}|100\rangle +\lambda _{2}|101\rangle ,  \label{b-ac-sd}
\end{equation}%
where $\lambda _{0}\lambda _{2}\neq 0$.

SD of the state belonging to C-AB SLOCC\ class is of the following form

\begin{equation}
|\psi ^{SD}\rangle =\lambda _{0}|000\rangle +\lambda _{1}e^{i\varphi
}|100\rangle +\lambda _{3}|110\rangle ,  \label{c-ab-sd}
\end{equation}%
where $\lambda _{0}\lambda _{3}\neq 0$.

\subsubsection{Theorem 2 is reduced for SD}

Theorem 2 can be reduced for SD as follows. For SD in Eq. (\ref{sd}), Eq. (%
\ref{cod-1}) becomes

\begin{equation}
\lambda _{0}^{2}+\lambda _{1}^{2}+\lambda _{3}^{2}=\lambda _{2}^{2}+\lambda
_{4}^{2}=1/2.  \label{SD-1}
\end{equation}

Eq. (\ref{cod-4}) becomes
\begin{equation}
\lambda _{1}\lambda _{2}=\lambda _{3}\lambda _{4}=0  \label{SD-2}
\end{equation}%
or
\begin{equation}
\phi =\pi \wedge \lambda _{1}\lambda _{2}=\lambda _{3}\lambda
_{4}\neq 0  \label{SD-3}
\end{equation}%
When $|\psi ^{SD}\rangle $ is chosen as the resource state, by directly
calculating $|\varphi \rangle _{a}|\psi ^{SD}\rangle $ Eqs. (\ref{SD-1}, \ref%
{SD-2}, \ref{SD-3}) can also be obtained.

From Theorem 2, we have the following corollary.

Corollary 1. SD of a state is suitable for PTP if and only if the SD of the
state\ satisfies the conditions in Eq. (\ref{SD-1}, \ref{SD-2}, \ref{SD-3}).

\subsubsection{PTP with GHZ SLOCC\ class}

For SD of GHZ\ SLOCC\ class, from Eqs. (\ref{ghz-sd}, \ref{SD-1}, \ref{SD-2}%
), obtain
\begin{eqnarray}
&&\frac{1}{\sqrt{2}}|000\rangle +\lambda _{2}|101\rangle +\lambda
_{4}|111\rangle ,\lambda _{2}\lambda _{4}\neq 0  \label{ghz-1} \\
&&\lambda _{0}|000\rangle +\lambda _{1}e^{i\phi }|100\rangle +\frac{1}{\sqrt{%
2}}|111\rangle ,\lambda _{0}\lambda _{1}\neq 0  \label{ghz-2} \\
&&\frac{1}{\sqrt{2}}(|000\rangle +|111\rangle ).  \label{ghz-3}
\end{eqnarray}

From Eqs. (\ref{ghz-sd}, \ref{SD-1}, \ref{SD-3}), obtain
\begin{equation}
\lambda _{0}|000\rangle -\lambda _{1}|100\rangle +\lambda _{2}|101\rangle
+\lambda _{3}|110\rangle +\lambda _{4}|111\rangle ,  \label{ghz-4}
\end{equation}%
where $\lambda _{0}^{2}+\lambda _{1}^{2}+\lambda _{3}^{2}=\lambda
_{2}^{2}+\lambda _{4}^{2}=1/2$ and $\lambda _{1}\lambda _{2}=\lambda
_{3}\lambda _{4}\neq 0$.

From \cite{Dli-jpa-20, Dli-qic-23}, one can see that the states from Eqs. (%
\ref{ghz-1} to \ref{ghz-4}) are LU inequivalent. Then, we can conclude the
following.

Result 1. SD of a state of GHZ SLOCC\ class is suitable for PTP if and only
if SD of the state\ is one of Eqs. (\ref{ghz-1} to \ref{ghz-4}). Via the LU
invariance for PTP, generally, a state of GHZ SLOCC\ class is suitable for
PTP if and only if the state is LU equivalent to one of the states in Eqs. (%
\ref{ghz-1} to \ref{ghz-4}).

Proposition 1. A state of GHZ SLOCC\ class has $S(\rho _{3})=\ln 2$, i.e.
has 1 ebit of shared entanglement between Alice and Bob, if and only if the
state is suitable for PTP.

Proof. Let $|\psi ^{SD}\rangle $ be SD of the state $|\psi \rangle $. There
are two steps for the proof.

Step 1. We show that $|\psi ^{SD}\rangle $ of GHZ SLOCC\ class has $S(\rho
_{3}(|\psi ^{SD}\rangle ))=\ln 2$ if and only if the $|\psi ^{SD}\rangle $
is suitable for PTP.

From \cite{Li-qic}, we know if $|\psi ^{SD}\rangle $ of GHZ SLOCC\ class has
$S(\rho _{3}(|\psi ^{SD}\rangle ))=\ln 2$, then the state must be one of the
states in Eqs. (\ref{ghz-1} to \ref{ghz-4}). By Result 1, the state is
suitable for PTP. Conversely, if $|\psi ^{SD}\rangle $ of of GHZ SLOCC\
class is suitable for PTP, by Result 1 the $|\psi ^{SD}\rangle $ is one of
Eqs. (\ref{ghz-1} to \ref{ghz-4}). A calculation yields that $S(\rho
_{3})=\ln 2$ for the states in Eq. (\ref{ghz-1} to \ref{ghz-4}). Ref. \cite%
{Li-qic}. Thus, it consumes 1 ebit of shared entanglement between Alice and
Bob .

Step 2. We next show that Proposition 1 holds generally via the LU
invariance for PTP and the LU invariance of von Neumann entropy.

Generally, assume that $S(\rho _{3}(|\psi \rangle ))=\ln 2$. Then, $S(\rho
_{3}(|\psi ^{SD}\rangle ))=\ln 2$ because $|\psi \rangle $ is LU equivalent
to $|\psi ^{SD}\rangle $ and the von Neumann entropy is LU invariant. By
Step 1, $|\psi ^{SD}\rangle $ is suitable for PTP. So, $|\psi \rangle $ is
suitable for PTP because PTP is LU invariant. Conversely, assume that $|\psi
\rangle $ is suitable for PTP. Then, $|\psi ^{SD}\rangle $ is suitable for
PTP because PTP is LU invariant. By Step 1, $S(\rho _{3}(|\psi ^{SD}\rangle
))=\ln 2$. Then, $S(\rho _{3}(|\psi \rangle ))=\ln 2$ because $|\psi \rangle
$ is LU equivalent to $|\psi ^{SD}\rangle $ and the von Neumann entropy is
LU invariant.

\subsubsection{PTP with W SLOCC\ class}

For SD of W\ SLOCC\ class, from Eq. (\ref{w-sd}), Eq. (\ref{SD-1}) becomes

\begin{equation}
\lambda _{0}^{2}+\lambda _{1}^{2}+\lambda _{3}^{2}=\lambda _{2}^{2}=1/2.
\label{w-2}
\end{equation}%
From the condition in Eq. (\ref{SD-2}), obtain $\lambda _{1}\lambda _{2}=0$.
Clearly, $\lambda _{2}\neq 0$ for W SLOCC\ class. Thus, obtain $\lambda
_{1}=0$. Since $\lambda _{4}=0$ for W SLOCC\ class, the case in Eq. (\ref%
{SD-3}) cannot happen.

Then, we can conclude the following.

Result 2. SD of W SLOCC\ class is suitable for PTP if and only if the SD\ is
of the following form.

\begin{equation}
|W_{pt}^{SD}\rangle =\lambda _{0}|000\rangle +\frac{\sqrt{2}}{2}|101\rangle
+\lambda _{3}|110\rangle ,  \label{w-1}
\end{equation}%
where $\lambda _{0}\lambda _{3}\neq 0$ and $\lambda _{0}^{2}+\lambda
_{3}^{2}=1/2$. Specially, when $\lambda _{0}=\lambda _{3}=\frac{1}{2}$, $%
|W_{pt}^{SD}\rangle $ becomes $\frac{1}{2}(|000\rangle +\sqrt{2}|101\rangle
+|110\rangle )$. Via the LU invariance for PTP, generally, a state of W
SLOCC\ class is suitable for \ PTP if and only if the state is LU equivalent
to the state which is of the form in Eq. (\ref{w-1}).

Ex. 4. Result 2 implies that $|W_{pt}^{SD}\rangle $ is a unique subclass of
W SLOCC\ class being suitable for PTP up to LU. One can see that SD of $%
\gamma |001\rangle +\beta |010\rangle +\alpha |100\rangle $ is $|\alpha
||000\rangle +|\gamma ||101\rangle +|\beta ||110\rangle $. Then, the SD of $%
|W_{n}\rangle _{123}$ is

\begin{equation}
|W_{n}^{SD}\rangle _{123}=\frac{1}{\sqrt{2}|\sqrt{n+1}|}|000\rangle +\frac{%
\sqrt{2}}{2}|101\rangle +\frac{|\sqrt{n}|}{\sqrt{2}|\sqrt{n+1}|}|110\rangle .
\label{wn-sd}
\end{equation}%
Clearly, $|W_{n}^{SD}\rangle _{123}$ is of the form $|W_{pt}^{SD}\rangle $
in Eq. (\ref{w-1}). Thus, $|W_{n}\rangle _{123}$ and its SD $%
|W_{n}^{SD}\rangle _{123}$ are suitable for \ PTP. One can say that $%
|W_{n}\rangle _{123}$ is a subclass of W SLOCC\ class, which is suitable for
\ PTP. Note that $n$ in $|W_{n}^{SD}\rangle _{123}$ is a natural number
while $\lambda _{0}$ and $\lambda _{3}$\ in $|W_{pt}^{SD}\rangle $ are
positive real numbers. So, the subclass $|W_{pt}^{SD}\rangle $ covers the
subclass $|W_{n}^{SD}\rangle _{123}$.

Proposition 2. A state of W SLOCC\ class has $S(\rho _{3})=\ln 2$, i.e. has
1 ebit of shared entanglement between Alice and Bob, if and only if the
state is suitable for PTP.

Proof. Ref. Proof for Proposition 1. Via the LU invariance for PTP and the
LU invariance of von Neumann entropy, we only need to show that SD of W
SLOCC\ class has $S(\rho _{3})=\ln 2$ if and only if the SD is suitable for
PTP.

Assume that SD of W SLOCC\ class has $S(\rho _{3})=\ln 2$, i.e. $\rho _{3}$
is the maximal mixed state, i.e. $\rho _{3}=tr_{12}\rho _{123}=(1/2)I$.
Then, from Eq. (\ref{w-sd}), a calculation yields $\lambda _{1}=0$ and $%
\lambda _{2}=\frac{1}{\sqrt{2}}$. Then, obtain the SD in Eq. (\ref{w-1}). By
Result 2, the SD is suitable for PTP. Conversely, if SD of W SLOCC\ class is
suitable for PTP, then by Result 2, it is $|W_{pt}^{SD}\rangle $ in Eq. (\ref%
{w-1}). A calculation yields $S(\rho _{3})=\ln 2$ for $|W_{pt}^{SD}\rangle $.

Ex. 5. For\ W state, its SD is $\frac{1}{\sqrt{3}}(|000\rangle +|101\rangle
+|110\rangle )$ which is not the form in Eq. (\ref{w-1}). For W state, $%
S(\rho _{3})\neq \ln 2$. Thus, W state and its SD are not suitable for PTP
by Proposition 2 and Result 2.

\subsubsection{PTP with A-BC SLOCC\ class}

For SD of A-BC\ SLOCC\ class, from Eq. (\ref{a-bc-sd}), Eq. (\ref{SD-1})
becomes

\begin{equation}
\lambda _{1}^{2}+\lambda _{3}^{2}=\lambda _{2}^{2}+\lambda _{4}^{2}=1/2.
\label{A-BC-2}
\end{equation}

From Eq. (\ref{SD-2}) and $\lambda _{1}\lambda _{4}e^{i\phi }-\lambda
_{2}\lambda _{3}\neq 0$, obtain
\begin{equation}
\lambda _{2}=\lambda _{3}=0\wedge \lambda _{1}\lambda _{4}\neq 0
\label{A-BC-3-1}
\end{equation}%
or
\begin{equation}
\lambda _{2}\lambda _{3}\neq 0\wedge \lambda _{1}=\lambda _{4}=0
\label{A-BC-3-2}
\end{equation}%
From Eqs. (\ref{SD-3}, \ref{A-BC-2}), obtain
\begin{equation}
\lambda _{1}=\lambda _{4}\wedge \lambda _{2}=\lambda _{3}  \label{A-BC-4}
\end{equation}

Thus, from Eqs. (\ref{A-BC-2}, \ref{A-BC-3-1}), obtain
\begin{equation}
\frac{1}{\sqrt{2}}(e^{i\phi }|100\rangle +|111\rangle ),  \label{A-BC-5}
\end{equation}

From Eqs. (\ref{A-BC-2}, \ref{A-BC-3-2}), obtain%
\begin{equation}
\frac{1}{\sqrt{2}}(|101\rangle +|110\rangle )  \label{A-BC-5-2}
\end{equation}%
From Eqs. (\ref{A-BC-2}, \ref{A-BC-4}), obtain
\begin{equation}
-\lambda _{1}|100\rangle +\lambda _{2}|101\rangle +\lambda _{2}|110\rangle
+\lambda _{1}|111\rangle ,  \label{A-BC-6}
\end{equation}%
where $\lambda _{1}^{2}+\lambda _{2}^{2}=1/2$ and $\lambda _{1}\lambda
_{2}\neq 0$.

Then, we can conclude the following.

Result 3. SD of A-BC SLOCC class is suitable for \ PTP if and only if the SD
is one of Eqs. (\ref{A-BC-5}, \ref{A-BC-5-2}, \ref{A-BC-6}). Via the LU
invariance for PTP, generally a state of A-BC SLOCC class is suitable for \
PTP if and only if the state is LU equivalent to one of Eqs. (\ref{A-BC-5}, %
\ref{A-BC-5-2}, \ref{A-BC-6}).

From \cite{Dli-jpa-20}, one can know the states in Eqs. (\ref{A-BC-5}, \ref%
{A-BC-5-2}, \ref{A-BC-6}) are LU equivalent. Therefore, Result 3 implies
that the LU class $\frac{1}{\sqrt{2}}(|101\rangle +|110\rangle )$\ is a
unique one which is suitable for PTP.

Proposition 3. A state of A-BC SLOCC\ class has $S(\rho _{3})=\ln 2$, i.e.
has 1 ebit of shared entanglement between Alice and Bob, if and only if the
state is suitable for PTP.

Proof. Ref. Proof for Proposition 1. Via the LU invariance for PTP and the
LU invariance of von Neumann entropy, we only need to show that SD of A-BC
SLOCC\ class has $S(\rho _{3})=\ln 2$ if and only if the SD is suitable for
PTP.

Assume that SD of A-BC SLOCC\ class \ has $S(\rho _{3})=\ln 2$, i.e. $\rho
_{3}=(1/2)I$.\ Then from Appendix B, the SD must be one of Eqs. (\ref{A-BC-5}%
, \ref{A-BC-5-2}, \ref{A-BC-6}). By Result 3, the SD is suitable for PTP.
Conversely, if SD of A-BC SLOCC\ class is suitable for PTP, then by Result
3, the SD must be one of Eqs. (\ref{A-BC-5}, \ref{A-BC-5-2}, \ref{A-BC-6}).
A calculation yields $S(\rho _{3})=\ln 2$ for the states in Eqs. (\ref%
{A-BC-5}, \ref{A-BC-5-2}, \ref{A-BC-6}).

\subsubsection{PTP with B-AC SLOCC\ class}

For SD of B-AC\ SLOCC\ class, from Eq. (\ref{b-ac-sd}), the condition in Eq.
(\ref{SD-3}) cannot happen for $|\psi ^{SD}\rangle $. Therefore, we only
consider the following cases. Eq. (\ref{SD-1}) becomes $\lambda
_{0}^{2}+\lambda _{1}^{2}=\lambda _{2}^{2}=1/2$. Eq. (\ref{SD-2}) becomes $%
\lambda _{1}=0$.

Then, we can conclude the following.

Result 4. SD of B-AC SLOCC class is suitable for \ PTP if and only if the SD
is uniquely the following state

\begin{equation}
\frac{1}{\sqrt{2}}(|000\rangle +|101\rangle ).  \label{B-AC-s}
\end{equation}

Via the LU invariance for PTP, generally \ a state of B-AC SLOCC class is
suitable for \ PTP if and only if the state is LU equivalent to the state in
Eq. (\ref{B-AC-s}).

Result 4 implies that the LU class $\frac{1}{\sqrt{2}}(|000\rangle
+|101\rangle )$ is a unique one which is suitable for PTP.

Proposition 4. A state of B-AC SLOCC\ class has $S(\rho _{3})=\ln 2$, i.e.
has 1 ebit of shared entanglement between Alice and Bob, if and only if the
state is suitable for \ PTP.

Proof. Ref. Proof for Proposition 1. Via the LU invariance for PTP and the
LU invariance of von Neumann entropy, we only need to show that SD of B-AC
SLOCC\ class has $S(\rho _{3})=\ln 2$ if and only if the SD is suitable for
\ PTP.

Assume that SD of B-AC SLOCC\ class \ has $S(\rho _{3})=\ln 2$, i.e. $\rho
_{3}=(1/2)I$. Then a calculation yields that the SD must be Eq. (\ref{B-AC-s}%
). By Result 4, the SD is suitable for PTP. The proof of other direction is
omitted.

\subsubsection{PTP with C-AB SLOCC\ class}

From Eq. (\ref{c-ab-sd}), it is easy to see that SD of C-AB\ SLOCC class
does not satisfy the condition in Eq. (\ref{SD-1}). So, no SD of C-AB\
SLOCC\ class is suitable for PTP. Therefore, via the LU invariance of PTP,
we can conclude the following:

Result 5. Any state of C-AB SLOCC class is not suitable for \ PTP.

For Eq. (\ref{c-ab-sd}), $\rho _{3}$ is the $diag(1,0)$ and then, $S(\rho
_{3})=0$ for C-AB SLOCC\ class. Therefore, from Result 5 the following
proposition 5 is still true.

Proposition 5. A state of C-AB SLOCC\ class has $S(\rho _{3})=\ln 2$, i.e.
has 1 ebit of shared entanglement between Alice and Bob, if and only if the
state is suitable for \ PTP.

Of course, the full separate states, i.e. all the states of A-B-C SLOCC\
class, are not suitable for PTP.

\subsubsection{General Results}

Propositions 1-5 imply the following theorem.

Theorem 3. A three-qubit state has $S(\rho _{3})=\ln 2$, i.e. has 1 ebit of
shared entanglement between Alice and Bob, if and only if the state is
suitable for \ PTP.

By Theorem 3, it is clear to see that GHZ state is suitable for PTP while W
state is not.

Remark 1. Note that the states in Eqs. (\ref{A-BC-5}, \ref{A-BC-5-2}, \ref%
{A-BC-6}, \ref{B-AC-s}) are separable and satisfy Corollary 1. Thus, these
separable states are suitable for PTP.\ Thus, clearly PTP does not require
genuine entanglement. But, it consumes 1 ebit of shared entanglement between
Alice and Bob .

\subsection{Bennett et al.'s teleportation is LU invariant while Nielsen and
Chuang's teleportation with CNOT and Hadmard gates is not LU invariant}

In \cite{Bennett}, the state $\frac{1}{\sqrt{2}}(|01\rangle -|10\rangle $
was used as the resource state for Bennett et al.'s teleportation. In
Appendix C, we show that a state is suitable for Bennett et al.'s
teleportation if and only if the state satisfies the following conditions

\begin{eqnarray}
|c_{0}|^{2}+|c_{2}|^{2} &=&|c_{1}|^{2}+|c_{3}|^{2}=\frac{1}{2},
\label{Bennett-1} \\
c_{0}c_{1}^{\ast }+c_{2}c_{3}^{\ast } &=&0  \label{Bennett-2}
\end{eqnarray}%
We also indicate in Appendix C that a state satisfies the conditions in Eqs.
(\ref{Bennett-1}, \ref{Bennett-2}) if and only if the state is one of the
following forms.

\begin{eqnarray}
|\psi _{1}\rangle _{12} &=&\frac{1}{\sqrt{2}}(e^{i\omega _{1}}|01\rangle
+e^{i\omega _{2}}|10\rangle ),  \label{state-1} \\
|\psi _{2}\rangle _{12} &=&\frac{1}{\sqrt{2}}(e^{i\omega _{0}}|00\rangle
+e^{i\omega _{3}}|11\rangle ),  \label{state-2} \\
|\psi _{3}\rangle _{12} &=&ke^{i\theta _{0}}|00\rangle +\ell e^{i\theta
_{1}}|01\rangle +\ell e^{i\theta _{2}}|10\rangle +ke^{i\theta
_{3}}|11\rangle ,  \label{state-3}
\end{eqnarray}%
where $\theta _{0}+\theta _{3}-(\theta _{1}+\theta _{2})=\pm \pi $ and $%
k^{2}+\ell ^{2}=1/2$ and $k\ell \neq 0$. Thus, $|\psi _{i}\rangle _{12}$, $%
i=1,2,3$, are all the suitable states for Bennett et al.'s teleportation.

It is easy to check that $|\psi _{i}\rangle _{12}$, $i=1,2,3$, are LU
equivalent to Bell state and have the maximal concurrence. Thus, Bennett et
al.'s teleportation is LU invariant.

In \cite{Nielsen}, Bell state $\frac{1}{\sqrt{2}}(|00\rangle +|11\rangle )$
was used as a resource state for teleportation, and CNOT, which is not local
operator, and Hadmard gates are applied to Alice's qubits \textquotedblleft
a\textquotedblright\ and 1. One can verify that the state $\frac{1}{2}%
(|00\rangle +i|01\rangle +i|10\rangle +|11\rangle )_{12}$ is LU equivalent
to Bell state. However, the state cannot be used as a resource state for
teleportation by the protocol in \cite{Nielsen}. It means that the
teleportation in \cite{Nielsen} is not LU invariant while Bennett et al.'s
teleportation is LU invariant.

One can check that the separable state $\frac{1}{2}(|00\rangle +i|01\rangle
-i|10\rangle +|11\rangle )_{12}$, which can be written as $\frac{1}{2}%
(|0\rangle -i|1\rangle )_{1}(|0\rangle +i|1\rangle )_{2}$, can be used as a
resource state for teleportation in \cite{Nielsen}. Thus, the teleportation
in \cite{Nielsen} does not require genuinely entangled states while Bennett
et al.'s teleportation \cite{Bennett} requires genuinely entangled states.

\subsection{A necessary and sufficient condition for a state of $n$ qubits
to be suitable for teleportation}

Assume that a state of $n$ qubit $|\psi \rangle _{1\cdots n}$ is suitable
for teleportation. Let Alice and Bob share a state of $n$ qubit state $|\psi
\rangle _{1\cdots n}$. Alice has qubits $1$, $\cdots ,$ $(n-1)$ while Bob
has qubit $n$. Alice has another qubit \textquotedblleft $a$%
\textquotedblright\ in the unknown state $|\varphi \rangle _{a}=(\alpha
|0\rangle _{a}+\beta |1\rangle _{a})$. Alice wants to send the unknown state
$|\varphi \rangle _{a}$ to Bob. She can perform a von Neumann joint
measurement on the qubits \textquotedblleft a\textquotedblright , 1, 2, $%
\cdots $, and $(n-1)$\ by using a set of orthogonal states\ and then convey
her result to Bob by sending 2-bit classical message. Then, Bob can convert
his state to the original unknown state\ by applying appropriate unitary
transformations. Thus, this completes the teleportation protocol. Ref. \cite%
{Bennett, Pankaj}.

Similarly, we can show that $|\psi \rangle _{12\cdots n}$ can be used as a
resource state for perfect teleportation if and only if it satisfies the
following Eqs. (\ref{n-q-1}, \ref{n-q-2}).

\begin{eqnarray}
\sum_{i=0}^{2^{n-1}-1}|c_{2i}|^{2} &=&\sum_{i=0}^{2^{n-1}-1}|c_{2i+1}|^{2}=%
\frac{1}{2}.  \label{n-q-1} \\
\sum_{i=0}^{2^{n-1}-1}c_{2i}c_{2i+1}^{\ast } &=&0  \label{n-q-2}
\end{eqnarray}

\section{PSDC-2}

Agrawal and Pati proposed PSDC-2 \cite{Pankaj}. They used GHZ state and
W-like state $|\eta _{1}^{+}\rangle _{123}=\frac{1}{2}(|010\rangle
+|001\rangle +\sqrt{2}|100\rangle )$ to demonstrate how their protocol works
as follows. For example, let Alice and Bob share GHZ state, where Alice has
the qubit 1 while Bob has the qubits 2 and 3. Alice can apply unitary
operators in $\{I,\sigma _{1},\sigma _{3},\sigma _{3}\sigma _{1}\}$ to her
qubit 1. Let
\begin{eqnarray}
A_{1} &=&I\otimes I\otimes I,  \label{loc-1} \\
A_{2} &=&\sigma _{3}\otimes I\otimes I,  \label{loc-2} \\
A_{3} &=&\sigma _{1}\otimes I\otimes I,  \label{loc-3} \\
A_{4} &=&\sigma _{3}\sigma _{1}\otimes I\otimes I.  \label{loc-4}
\end{eqnarray}%
That is, the local unitary operators $A_{i}$, $i=1,2,3,4$, are applied to
the state GHZ.

It is necessary for these operators to convert the original state to a set
of four orthogonal states $A_{i}|GHZ\rangle $, $i=1,2,3,4$. Then, Alice
sends her qubit to Bob, and then, Bob can perform von Neumann measurement on
qubits 1, 2 and 3 via these orthogonal states. Thus, Bob can determine what
unitary operator Alice has applied. Thus, Alice can transmit 2-bit classical
message by sending one qubit.

\subsection{PSDC-2 is LU invariant}

In this subsection, we show that PSDC-2 is\ LU invariant. It is enough to
show that $|\psi \rangle $ is suitable for PSDC-2 if and only if its SD $%
|\psi ^{SD}\rangle $ is. It is also enough to show that when $|\psi \rangle $
is used as a resource state, any two of $A_{i}|\psi \rangle $, $i=1,2,3,4$,
are orthogonal if and only if when its SD is used as a resource state, any
two of $A_{i}|\psi ^{SD}\rangle $, $i=1,2,3,4$, are orthogonal. The proof is
put in Appendix D. Via the invariance, it is easy to describe what states
are suitable for PSDC-2.

\subsection{A necessary and sufficient condition for PSDC-2}

Agrawal and Pati indicated that using W state, Alice cannot transmit 2-bit
classical information by sending one qubit. In this subsection, we find all
the states of three qubits which are suitable for PSDC-2.

Let\ $|T^{+}\rangle =\sum_{i=0}^{7}c_{i}|i\rangle $ be a resource state for
PSDC-2. Alice applies the unitary operators to her qubit, then obtain $%
A_{1}|T^{+}\rangle =|T^{+}\rangle ,$ $A_{2}|T^{+}\rangle =|T^{-}\rangle ,$ $%
A_{3}|T^{+}\rangle =|H^{+}\rangle ,$ $A_{4}|T^{+}\rangle =|H^{-}\rangle $.
Then, Alice sends her qubit 1 to Bob. After receiving from Alice, for
perfect PSDC-2, Bob performs von Neumann measurement on qubits 1, 2, and 3.
To this end, it needs to guarantee that $|T^{\pm }\rangle $ and $|H^{\pm
}\rangle $ are a set of orthogonal states.

A calculation yields $\langle H^{+}|H^{-}\rangle =-\langle
T^{+}|T^{-}\rangle $ and $\langle T^{+}|T^{-}\rangle
=\sum_{i=0}^{3}|c_{i}|^{2}-\sum_{i=0}^{3}|c_{i+4}|^{2}$. From that $%
\sum_{i=0}^{7}|c_{i}|^{2}=1$, then $\langle T^{+}|T^{-}\rangle =0$ if and
only if

\begin{equation}
\sum_{i=0}^{3}|c_{i}|^{2}=\sum_{i=4}^{7}|c_{i}|^{2}=\frac{1}{2}
\label{1-qubit-1}
\end{equation}

One can check that $\langle T^{-}|H^{-}\rangle =\langle T^{+}|H^{+}\rangle $
and $\langle T^{-}|H^{+}\rangle =\langle T^{+}|H^{-}\rangle $. Then, $%
\langle T^{+}|H^{+}\rangle =\langle T^{+}|H^{-}\rangle =0$ if and only if
\begin{equation}
c_{0}^{\ast }c_{4}+c_{1}^{\ast }c_{5}+c_{2}^{\ast }c_{6}+c_{3}^{\ast }c_{7}=0
\label{1-qubit-2}
\end{equation}

Thus, we can conclude the following.

Theorem 4. $\sum_{i=0}^{7}c_{i}|i\rangle $ is suitable for PSDC-2 if and
only if the state satisfies Eqs. (\ref{1-qubit-1}, \ref{1-qubit-2}).

The detailed proof is put in Appendix E.

Ex. 6. Let $|C\rangle =(1/2)(|000\rangle +|011\rangle +|100\rangle
-|111\rangle )$. Then, one can check that $|C\rangle =M\otimes I\otimes
I|GHZ\rangle $, where $M$ is the hadmard matrix. Thus, $|C\rangle $ is LU
equivalent to GHZ. It is easy to verify that $|C\rangle $, $|G\rangle $ in
Ex. 2, GHZ state, and the graph states in Eqs. (\ref{graph-1}, \ref{graph-2}%
) satisfy Theorem 4.

Ex. 7. The W-like state $\gamma |001\rangle +\beta |010\rangle +\frac{1}{%
\sqrt{2}}|100\rangle $, where $\beta $ and $\gamma $\ are complex and $%
|\beta |^{2}+|\gamma |^{2}=1/2$,\ and its SD $\frac{1}{\sqrt{2}}|000\rangle
+|\gamma ||101\rangle +|\beta ||110\rangle $\ satisfy the conditions of
Theorem 4.

Ex. 8. One can check that $|W_{n}\rangle _{123}$ and its SD $%
|W_{n}^{SD}\rangle _{123}$ do not satisfy Eq. (\ref{1-qubit-1}). Therefore, $%
|W_{n}\rangle _{123}$ and $|W_{n}^{SD}\rangle _{123}$\ are not suitable for
PSDC-2. If Alice has qubit 3, then $|W_{n}\rangle _{123}$ may be suitable
for PSDC-2 \cite{LLI}.

Ex. 9. Let $|\Phi \rangle =\frac{1}{\sqrt{6}}(|000\rangle +|010\rangle
+|100\rangle +|101\rangle -|110\rangle +|111\rangle )$. $|\Phi \rangle $ is
entangled, but its 3-tangle vanishes, so the state belongs to W SLOCC\
class. For $|\Phi \rangle $, a calculation yields its SD $|\Phi ^{SD}\rangle
=\frac{1}{\sqrt{3}}(|000\rangle +|101\rangle +|110\rangle )$. $|\Phi \rangle
$ and its SD $|\Phi ^{SD}\rangle $ do not satisfy Eq. (\ref{1-qubit-1}). One
can also check that W state and its SD $\frac{1}{\sqrt{3}}(|000\rangle
+|101\rangle +|110\rangle )$ are not suitable for PSDC-2.

\subsection{PSDC-2 with SD}

In this subsection, we want to find all SDs which are suitable for PSDC-2.
Then, via the LU invariance of PSDC-2, we get all the states which are
suitable for PSDC-2.

For $|\psi ^{SD}\rangle $ in Eq, (\ref{sd}), Eq. (\ref{1-qubit-1}) becomes $%
\lambda _{0}^{2}=\frac{1}{2}$. Eq. (\ref{1-qubit-2}) becomes $\lambda
_{0}\lambda _{1}e^{i\phi }=0$. Thus, obtain $\lambda _{1}=0$.

From Theorem 4, we have the following corollary.

Corollary 2. Any SD is suitable for \ PSDC-2 if and only if it is of the form

\begin{equation}
\frac{1}{\sqrt{2}}|000\rangle +\lambda _{2}|101\rangle +\lambda
_{3}|110\rangle +\lambda _{4}|111\rangle ,  \label{sd-1}
\end{equation}%
where $\lambda _{2}^{2}+\lambda _{3}^{2}+\lambda _{4}^{2}=1/2.$

\subsubsection{PSDC-2 with GHZ SLOCC\ class}

From Corollary 2 and Eq. (\ref{ghz-sd}), obtain the following.

Result 6. SD of GHZ SLOCC\ class is suitable for \ PSDC-2 if and only if it
is of the following form
\begin{equation}
\frac{1}{\sqrt{2}}|000\rangle +\lambda _{2}|101\rangle +\lambda
_{3}|110\rangle +\lambda _{4}|111\rangle ,  \label{ghz-sdc}
\end{equation}%
where $\lambda _{2}^{2}+\lambda _{3}^{2}+\lambda _{4}^{2}=1/2$ and $\lambda
_{4}\neq 0$. Via the LU invariance of PSDC-2, generally, a state of GHZ
SLOCC\ class is suitable for PSDC-2 if and only if it is LU equivalent to
the state in Eq. (\ref{ghz-sdc}).

Proposition 6. A state of GHZ SLOCC\ class has $S(\rho _{1})=\ln 2$, i.e.
has 1 ebit of shared entanglement between Alice and Bob, if and only if the
state is suitable for \ PSDC-2.

Proof. Ref. Proof for Proposition 1. Via the LU invariance for PSDC-2 and
the LU invariance of von Neumann entropy, we only need to show that SD of
GHZ SLOCC\ class has $S(\rho _{1})=\ln 2$ if and only if the SD is suitable
for\ PSDC-2.

Assume that SD of GHZ SLOCC\ class\ has $S(\rho _{1})=\ln 2$. Then from\
\cite{Li-qic}, the SD must be the state in Eq. (\ref{ghz-sdc}). By Result 6,
the SD is suitable for PSDC-2. Conversely, if SD of GHZ SLOCC\ class is
suitable for PTP, then by Result 6, the SD must be Eq. (\ref{ghz-sdc}). A
calculation yields $S(\rho _{1})=\ln 2$ for the state in Eq. (\ref{ghz-sdc}).

\subsubsection{PSDC-2 with W SLOCC\ class}

From Corollary 2 and Eq. (\ref{w-sd}), obtain the following.

Result 7. SD of W\ SLOCC\ class is suitable for\ PSDC-2 if and only if it is
of the form
\begin{equation}
|W_{sdc}^{SD}\rangle =\frac{1}{\sqrt{2}}|000\rangle +\lambda _{2}|101\rangle
+\lambda _{3}|110\rangle ,  \label{w-sdc-sd}
\end{equation}%
where $\lambda _{2}^{2}+\lambda _{3}^{2}=1/2$ and $\lambda _{2}\lambda
_{3}\neq 0$. Via the LU invariance of PSDC-2, generally, a state of W SLOCC\
class is suitable for PSDC-2 if and only if it is LU equivalent to the state
of the form $|W_{sdc}^{SD}\rangle $ in Eq. (\ref{w-sdc-sd}).

Result 7 implies that $|W_{sdc}^{SD}\rangle $ is a unique subclass of W
SLOCC\ class\ up to LU which is suitable for PSDC-2. Thus, we find all the
states of W SLOCC\ class which can be suitable for PSDC-2.

Proposition 7. A state of W SLOCC\ class has $S(\rho _{1})=\ln 2$, i.e. has
1 ebit of shared entanglement between Alice and Bob, if and only if the
state is suitable for\ PSDC-2.

Proof. Ref. Proofs for Propositions 1 and 6. Via the LU invariance for
PSDC-2 and the LU invariance of von Neumann entropy, we only need to show
that SD of W SLOCC\ class has $S(\rho _{1})=\ln 2$ if and only if the SD is
suitable for\ PSDC-2.

Assume that SD of W SLOCC\ class\ has $S(\rho _{1})=\ln 2$, i.e. $\rho
_{1}=tr_{23}\rho _{123}=(1/2)I$. Then, from Eq. (\ref{w-sd}), a calculation
yields $\lambda _{0}=\frac{1}{\sqrt{2}}$ and $\lambda _{1}=0$. Then, the SD
must be the state $|W_{sdc}^{SD}\rangle $. By Result 7, the SD is suitable
for PSDC-2. Conversely, if SD of W SLOCC\ class is suitable for PSDC-2, then
by Result 7, the SD must be $|W_{sdc}^{SD}\rangle $ in Eq. (\ref{w-sdc-sd}).
A calculation yields $S(\rho _{1})=\ln 2$ for $|W_{sdc}^{SD}\rangle $ in Eq.
(\ref{w-sdc-sd}).

\subsubsection{PSDC-2 with A-BC SLOCC\ class}

The states in Eq. (\ref{a-bc-sd}) do not satisfy Corollary 2. So, no SD of
A-BC\ SLOCC\ class is suitable for PSDC-2. Therefore, via the LU invariance
of PSDC-2, we can conclude the following.

Result 8. Any state of A-BC\ SLOCC\ class is not suitable for PSDC-2.

For the state in Eq. (\ref{a-bc-sd}), a calculation yields $\rho
_{1}=diag(1,0)$ and then, $S(\rho _{1})=0$ for A-BC SLOCC\ class. Therefore,
from Result 8 the following proposition 8 is still true.

For example, the states in Eqs. (\ref{A-BC-5}, \ref{A-BC-5-2}, \ref{A-BC-6})
are not suitable for PSDC-2.

Proposition 8. A state of A-BC SLOCC\ class has $S(\rho _{1})=\ln 2$, i.e.
has 1 ebit of shared entanglement between Alice and Bob, if and only if the
state is suitable for\ PSDC-2.

\subsubsection{PSDC-2 with B-AC SLOCC\ class}

From Corollary 2 and Eq. (\ref{b-ac-sd}), obtain the following.

Result 9. SD of B-AC class is suitable for PSC-2 if and only if it is$\frac{1%
}{\sqrt{2}}(|000\rangle +|101\rangle )$ (a separable state). Via the LU
invariance of PSDC-2, generally, a state of B-AC class is suitable for
PSDC-2 if and only if it is LU equivalent to $\frac{1}{\sqrt{2}}(|000\rangle
+|101\rangle )$.

Result 9 implies that the LU class $\frac{1}{\sqrt{2}}(|000\rangle
+|101\rangle )$ is a unique one which is suitable for PSDC-2.

Proposition 9. A state of B-AC SLOCC\ class has $S(\rho _{1})=\ln 2$, i.e.
has 1 ebit of shared entanglement between Alice and Bob, if and only if the
state is suitable for\ PSDC-2.

Proof. Ref. Proofs for Propositions 1 and 6. Via the LU invariance for
PSDC-2 and the LU invariance of von Neumann entropy, we only need to show
that SD of B-AC SLOCC\ class has $S(\rho _{1})=\ln 2$ if and only if the SD
is suitable for\ PSDC-2. The proof is omitted.

\subsubsection{PSDC-2 with C-AB SLOCC\ class}

From Corollary 2 and Eq. (\ref{c-ab-sd}), obtain the following.

Result 10. SD of C-AB class is suitable for\ PSDC-2 if and only if it is$%
\frac{1}{\sqrt{2}}(|000\rangle +|110\rangle )$ (which is a separable state).
Via the LU invariance of PSDC-2, generally, a state of C-AB class is
suitable for PSDC-2 if and only if it is LU equivalent to $\frac{1}{\sqrt{2}}%
(|000\rangle +|110\rangle )$.

Result 10 implies that the LU class $\frac{1}{\sqrt{2}}(|000\rangle
+|110\rangle )$ is a unique one which is suitable for PSDC-2.

Proposition 10. A state of C-AB SLOCC\ class has $S(\rho _{1})=\ln 2$, i.e.
has 1 ebit of shared entanglement between Alice and Bob , if and only if the
state is suitable for\ PSDC-2.

Proof. Ref. Proofs for Propositions 1 and 6. Via the LU invariance for
PSDC-2 and the LU invariance of von Neumann entropy, we only need to show
that SD of C-AB SLOCC\ class has $S(\rho _{1})=\ln 2$ if and only if the SD
is suitable for\ PSDC-2. The proof is omitted.

\subsubsection{General results}

Propositions 6-10 implies the following theorem.

Theorem 5. A three-qubit state has $S(\rho _{1})=\ln 2$, i.e. has 1 ebit of
shared entanglement between Alice and Bob, if and only if the state is
suitable for\ PSDC-2.

By Theorem 5, clearly GHZ state is suitable for PSCD-2 while W state is not.

Remark 2. PSDC-2 does not require genuine entanglement. But, it consumes 1
ebit of shared entanglement between Alice and Bob.

Agrawal and Pati used genuinely entangled states \ GHZ and $|\eta
_{1}^{+}\rangle _{123}$\ to demonstrate how PSDC-2 works. We use the
following examples to show that some separable states are also suitable for
PSDC-2.

$\frac{1}{\sqrt{2}}(|000\rangle +|101\rangle )$ of B-AC\ SLOCC\ class and $%
\frac{1}{\sqrt{2}}(|000\rangle +|110\rangle )$ of C-AB\ SLOCC\ class are
separable. It is clear that they satisfy the conditions of Theorem 4,
therefore they are suitable for PSDC-2. For another example, let $|\varpi
^{+}\rangle =\frac{1}{\sqrt{2}}(|011\rangle +|110\rangle )$, which is
separable and LU equivalent to $\frac{1}{\sqrt{2}}(|000\rangle +|101\rangle
) $. $|\varpi ^{+}\rangle $ satisfies the conditions of Theorem 4. It means
that $|\varpi ^{+}\rangle $ is suitable for PSDC-2. A calculation can also
verify that $A_{i}|\varpi ^{+}\rangle $, $i=1,2,3,4$, are a set of four
orthogonal states $\frac{1}{\sqrt{2}}(|011\rangle _{123}\pm |110\rangle
_{123})$ and $\frac{1}{\sqrt{2}}(|010\rangle _{123}\pm |111\rangle _{123})$.

\section{PSDC-3}

Agrawal and Pati proposed PSDC-3 \cite{Pankaj}. They used GHZ state to
demonstrate how PSDC-3 works as follows. Let Alice and Bob share GHZ state.
Alice has qubits 1 and 2, while Bob has qubit 3. Alice can apply local
unitary transformations on her qubits. Besides the local unitary operators $%
A_{i}$, $i=1,2,3,4$ in section 3, she also applies the following local
unitary operators on her qubits.
\begin{eqnarray}
A_{5} &=&I\otimes \sigma _{1}\otimes I,  \label{loc-5} \\
A_{6} &=&I\otimes \sigma _{3}\sigma _{1}\otimes I,  \label{loc-6} \\
A_{7} &=&\sigma _{1}\otimes \sigma _{1}\otimes I,  \label{loc-7} \\
A_{8} &=&\sigma _{1}\otimes \sigma _{3}\sigma _{1}\otimes I  \label{loc-8}
\end{eqnarray}

Thus, obtain a set of the orthogonal states $A_{i}|GHZ\rangle $, $i=1$ to $8$%
. After Alice sends qubits 1 and 2 to Bob, Bob can make von Neumann
measurement using the orthogonal set \{$A_{i}|GHZ\rangle $, $i=1$ to $8$\}.
That is, Bob can perfectly distinguish what operation Alice has applied.
Therefore, he can recover 3-bit classical information.

\subsection{A necessary and sufficient condition for PSDC-3}

In this section, we want to find all the states which are suitable for
PSDC-3. From Appendix E, we obtain the following.

Theorem 6. Let $|\psi \rangle =\sum_{i=0}^{7}c_{i}|i\rangle $ be any state
of three qubits. Then, $|\psi \rangle $ is suitable for PSDC-3 if and only
if it satisfies the following conditions
\begin{eqnarray}
|c_{0}|^{2}+|c_{1}|^{2} &=&|c_{6}|^{2}+|c_{7}|^{2}  \label{x3} \\
|c_{2}|^{2}+|c_{3}|^{2} &=&|c_{4}|^{2}+|c_{5}|^{2}  \label{x4} \\
\mathrm{Re}(c_{2}^{\ast }c_{6}+c_{3}^{\ast }c_{7}) &=&0  \label{a1} \\
c_{0}^{\ast }c_{4}+c_{1}^{\ast }c_{5}+c_{2}^{\ast }c_{6}+c_{3}^{\ast }c_{7}
&=&0  \label{a3} \\
c_{0}^{\ast }c_{2}+c_{1}^{\ast }c_{3} &=&0  \label{u1} \\
c_{4}^{\ast }c_{6}+c_{5}^{\ast }c_{7} &=&0  \label{u2} \\
c_{2}c_{4}^{\ast }+c_{3}c_{5}^{\ast } &=&0  \label{t1} \\
c_{0}^{\ast }c_{6}+c_{1}^{\ast }c_{7} &=&0  \label{t2}
\end{eqnarray}

Ex. 10. Let $|\kappa _{1}\rangle =(1/2)(|010\rangle +|011\rangle
+|100\rangle -|101\rangle )$. It is easy to verify that $|\kappa _{1}\rangle
=I\otimes \sigma _{1}\otimes M|GHZ\rangle $, where $M$ is the hadmard
matrix. For $|\kappa _{1}\rangle $, 3-tangle is 1.\ Let $|\kappa _{2}\rangle
=\frac{1}{\sqrt{2}}(|010\rangle +|101\rangle )$ and $|\kappa _{3}\rangle =%
\frac{1}{\sqrt{2}}(|011\rangle +|100\rangle )$. Clearly, $|\kappa
_{i}\rangle $, $i=1,2,3$, are LU equivalent to the state GHZ and satisfy the
conditions of Theorem 6.

\subsection{PSDC-3 is not LU invariant.}

From Ex.6, $|C\rangle $ is LU equivalent to GHZ. As mentioned before, $%
|C\rangle $ is suitable for PSDC-2. But $|C\rangle $ is not suitable for
PSDC-3 because it does not satisfy Eq. (\ref{a1}). We can also show that $%
|C\rangle $ is not suitable for PSDC-3\ by checking that $A_{5}|C\rangle $
and $A_{8}|C\rangle $ are not orthogonal. But, GHZ is suitable for PSDC-3
\cite{Pankaj}. Therefore, PSDC-3 is not LU invariant.

\subsection{PSDC-3 with SD}

Via Theorem 6, we can derive the conditions for $|\psi ^{SD}\rangle $ in Eq.
(\ref{sd})\ for PSDC-3. Via Eq. (\ref{x4}), $\lambda _{1}=\lambda _{2}=0$.
Via Eq. (\ref{x3}), obtain $\lambda _{0}^{2}=\lambda _{3}^{2}+\lambda
_{4}^{2}=1/2$ and then $\lambda _{0}=1/\sqrt{2}$. Via Eq. (\ref{t2}), obtain
$\lambda _{0}\lambda _{3}=0$. And then, $\lambda _{3}=0$ because $\lambda
_{0}=1/\sqrt{2}$. Then, $\lambda _{4}=1/\sqrt{2}$. Therefore, $|\psi
^{SD}\rangle $ satisfies the conditions of Theorem 6 if and only if it is
just the GHZ state $\frac{1}{\sqrt{2}}(|000\rangle +|111\rangle )$.

We conclude the following.

Corollary 3. GHZ state $\frac{1}{\sqrt{2}}(|000\rangle +|111\rangle )$ is a
unique SD which is suitable for PSDC-3.

Agrawal and Pati suggested to study if there are subclasses of W SLOCC\
class which are suitable for PSDC-3. Clearly, SD of W SLOCC\ class in Eq. (%
\ref{w-sd}) does not satisfy the conditions of Theorem 6. We still cannot
determine if some states of W SLOCC\ class satisfy the conditions of Theorem
6 because PSDC-3 is not LU invariant.

One can check that W-like states $\alpha |100\rangle +\beta |010\rangle
+\gamma |001\rangle $, $|W_{n}\rangle _{123}$ in \cite{Pankaj},\ and Dicke
states $a|110\rangle +b|101\rangle +c|011\rangle $\ do not satisfy Eq. (\ref%
{x3}). While W-like states, $|W_{n}\rangle _{123}$ in \cite{Pankaj}, and
Dicke states belong to W SLOCC\ class.

So far, Agrawal and Pati's suggestion still remains an open question. By
solving Eqs. (\ref{x3}-\ref{t2}) for PSDC-3, we show that any state of the
SLOCC\ classes W is not suitable for PSDC-3 and obtain the following theorem.

Theorem 7.

(1). Any state of the SLOCC\ class W is not suitable for PSDC-3.

(2). Any state of the SLOCC\ classes A-BC, B-AC,C-AB, and A-B-C is not
suitable for PSDC-3.

(3). Let $|\psi \rangle _{123}=\sum_{i=0}^{7}c_{i}|i\rangle _{123}$ be any
state of GHZ SLOCC\ class.

(3.1). When $c_{i}\neq 0$, $i=0,...,7$, it is unknown if Eqs. (\ref{x3}-\ref%
{t2}) has a solution.

(3.2). When some of $c_{0}$ to $c_{7}$ vanish, we find all states of GHZ
SLOCC\ class, which are suitable for PSDC-3. These states are $|F_{i}\rangle
$, $i=0,1,2,3$, and $|\pi _{i}\rangle ,$ $i=1,2,3,4$.%
\begin{eqnarray}
|F_{0}\rangle &=&(1/\sqrt{2})(e^{i\theta _{0}}|000\rangle +e^{i\theta
_{7}}|111\rangle ), \\
|F_{1}\rangle &=&(1/\sqrt{2})(e^{i\theta _{1}}|001\rangle +e^{i\theta
_{6}}|110\rangle ), \\
|F_{2}\rangle &=&(1/\sqrt{2})(e^{i\theta _{2}}|010\rangle +e^{i\theta
_{5}}|101\rangle ), \\
|F_{3}\rangle &=&(1/\sqrt{2})(e^{i\theta _{3}}|011\rangle +e^{i\theta
_{4}}|100\rangle ), \\
|\pi _{1}\rangle &=&c_{2}|010\rangle +c_{3}|011\rangle +c_{4}|100\rangle
+c_{5}|101\rangle , \\
|\pi _{2}\rangle &=&c_{0}|000\rangle +c_{1}|001\rangle +c_{6}|110\rangle
+c_{7}|111\rangle , \\
|\pi _{3}\rangle &=&c_{1}|001\rangle +c_{2}|010\rangle +c_{5}|101\rangle
+c_{6}|110\rangle , \\
|\pi _{4}\rangle &=&c_{0}|000\rangle +c_{3}|011\rangle +c_{4}|100\rangle
+c_{7}|111\rangle .
\end{eqnarray}

The proof is put in Appendix F. For the conditions for these states, see
Appendix F.

\section{Comparison}

Difference 1.

For PSDC-2, Alice has qubit 1 while Bob has qubits 2 and 3, and Alice
applies unitary operators in $\{I,\sigma _{1},\sigma _{3},\sigma _{3}\sigma
_{1}\}$ to her qubit 1. We indicate that $|W_{n}\rangle _{123}$ in \cite%
{Pankaj}\ \ is not suitable for PSDC-2. Li and Qiu also studied superdense
coding to transmit 2-bit classical information by sending one qubit. For
their protocol, Alice has qubit 3 while Bob has qubits 1 and 2. Alice
applies $\{I,\sigma _{1},\sigma _{3},-i\sigma _{2}\}$ on her qubit. They
indicated that $|W_{n}\rangle _{123}$ is suitable for their protocol.

One can see that for PSDC-2, Alice has qubit 1 while for Li and Qiu's
protocol, Alice has qubit 3. Therefore, PSDC-2 is different from Li and
Qiu's protocol. Even one can also study a protocol different from the above
two ones, where Alice has qubit 2 and Bob has qubits 1 and 3.

Difference 2.

For PSDC-3, we point out that $|W_{n}\rangle _{123}$\ is not suitable for
PSDC-3. Li and Qiu proposed their protocol for superdense coding to transmit
3-bit classical information by sending two qubits, where Alice has qubits 1
and 2 while Bob has qubit 3. They indicated $|W_{n}\rangle _{123}$ is
suitable for their protocol. Note that they mentioned that Alice needs to
choose operators $U_{x}\in u_{12}U_{12}^{\dag }$ on her two qubits to use
the two-qubit unitary operation $U_{12}$ and a one-qubit unitary operation $%
I_{3}$ to convert $|GHZ\rangle _{123}$ to $|W_{1}\rangle _{123}$. That is, $%
U_{12}\otimes I_{3}|GHZ\rangle _{123}=|W_{1}\rangle _{123}$ \cite{LLI}.
Clearly, the two-qubit unitary operation $U_{12}$ is not a local unitary
operator. While for PSDC-3, Alice applies local unitary operators in Eqs. (%
\ref{loc-1} to \ref{loc-4}, \ref{loc-5} to \ref{loc-8}) to her qubits. This
is their difference between them.

Difference 3.

In this paper, we find all separable states which are suitable for PTP and
PSDC-2. Thus, PTP and PSDC-2 do not require genuine entanglement. While
previous papers indicated that use of nonmaximally entangled resources leads
to probabilistic protocols for teleportation and probabilistic superdense
coding \cite{WL, GG, SM, AP-05, Singh}.

Difference 4. We propose a concept about LU invariance of a protocol, which
has not appeared in previous papers.

Difference 5. Agrawal and Pati suggested to study if there are subclasses of
W SLOCC class which are suitable for PSDC-3. So far, it still remains an
unsolved question. In this paper, we show that no subclass of W SLOCC class
is suitable for PSDC-3. Thus, we solve the unsolved question.

Difference 6. We find all suitable states for PTP, PSDC-2, and PSDC-3.

Difference 7. We prove that a state is suitable for PTP and PSDC-2 if and
only if it has 1 ebit of shared entanglement between Alice and Bob,
respectively.

\section{Conclusion}

In this paper, we show that PTP and PSDC-2 are LU invariant and indicate
that PSDC-3 is not LU invariant. Via the LU invariance of PTP and PSDC-2, we
prove that a state is suitable for PTP and PSDC-2 if and only if it has 1
ebit of shared entanglement between Alice and Bob, respectively. Thus, one
can determine if a state is suitable for PTP or PSDC-2 by checking if the
state has 1 ebit of shared entanglement between Alice and Bob.

We propose a necessary and sufficient condition for a state to be suitable
for PTP, PSDC-2, and PSDC-3, respectively. The condition only requires
additions and multiplications for the coefficients of states. Via the LU
invariance of PTP and PSDC-2, we find all the states which are suitable for
PTP and PSDC-2. Interestingly, we find all separable states which are
suitable for PTP and PSCD-2. It implies that PTP and PSDC-2 do not require
genuine entanglement. However, they do consume 1 ebit of shared entanglement.

Previous articles have explored if there are other subclasses of W SLOCC\
class which are suitable for PTP and PSCD-2. Via the LU invariance of PTP,
we show that there is a unique subclass $|W_{pt}^{SD}\rangle $\ of W SLOCC
class which is suitable for PTP up to LU. Via the LU invariance of PSDC-2,
we find a unique subclass $|W_{sdc}^{SD}\rangle $\ \ of W SLOCC\ class which
is suitable for PSDC-2 up to LU.

Agrawal and Pati suggested to study if there are subclasses of W SLOCC class
which are suitable for PSDC-3. So far, it still remains an unsolved
question. By solving Eqs. (\ref{x3}-\ref{t2}) for PSDC-3, we show that any
state of the SLOCC\ classes W is not suitable for PSDC-3.

\subsection{Statements and declarations}

No financial interests, no competing interests, no financial supports.

\subsection{A data availability statement}

It includes all data in the main text.

Acknowledgements

Thank the reviewers for their useful comments and Julia Huang (of Stanford
University) for changing English.

\section{Appendix A. PTP is LU invariant}

\setcounter{equation}{0} \renewcommand{\theequation}{A\arabic{equation}} Let
$|\psi \rangle _{123}=\sum_{i=0}^{7}c_{i}|i\rangle _{123}$. It is well known
that any state $|\psi \rangle _{123}$ can be transformed into SD $|\psi
^{SD}\rangle _{123}$ in Eq. (\ref{sd}) under LU. That is,
\begin{equation}
|\psi ^{SD}\rangle _{123}=U_{1}\otimes U_{2}\otimes U_{3}|\psi \rangle
_{123}.
\end{equation}

Let $U_{i}^{\dag }$ be the Hermitian transpose of $U_{i}$, $i=1,2,3$. Then,
conversely obtain

\begin{equation}
|\psi \rangle _{123}=U_{1}^{\dag }\otimes U_{2}^{\dag }\otimes U_{3}^{\dag
}|\psi ^{SD}\rangle _{123}.
\end{equation}

Let $I_{a}$ be the identity and $|\varphi \rangle _{a}=\alpha |0\rangle
_{a}+\beta |1\rangle _{a}$. Then,
\begin{eqnarray}
|\varphi \rangle _{a}|\psi \rangle _{123} &=&(I_{a}\otimes U_{1}^{\dag
}\otimes U_{2}^{\dag }\otimes U_{3}^{\dag })(I_{a}\otimes U_{1}\otimes
U_{2}\otimes U_{3})|\varphi \rangle _{a}|\psi \rangle _{123} \\
&=&(I_{a}\otimes U_{1}^{\dag }\otimes U_{2}^{\dag }\otimes U_{3}^{\dag
})|\varphi \rangle _{a}|\psi ^{SD}\rangle _{123}.
\end{eqnarray}

Proof. Assume that $|\psi \rangle _{123}$ is suitable for perfect
teleportation. Then, by Agrawal-Pati protocol, we can obtain
\begin{eqnarray}
&&|\varphi \rangle _{a}|\psi \rangle _{123} \\
&=&\frac{1}{2}[|\chi ^{+}\rangle _{a12}(I|\varphi \rangle _{3})+|\chi
^{-}\rangle _{a12}(\sigma _{3}|\varphi \rangle _{3}) \\
&&+|\varsigma ^{+}\rangle _{a12}(\sigma _{1}|\varphi \rangle
_{3})+|\varsigma ^{-}\rangle _{a12}(\sigma _{3}\sigma _{1}|\varphi \rangle
_{3})],
\end{eqnarray}%
where $|\chi ^{\pm }\rangle _{a12}$ and $|\varsigma ^{\pm }\rangle _{a12}$
are a set of orthogonal states.

Then,

\begin{eqnarray}
&&|\varphi \rangle _{a}|\psi ^{SD}\rangle _{123}  \nonumber \\
&=&I_{a}\otimes U_{1}\otimes U_{2}\otimes U_{3}|\varphi \rangle _{a}|\psi
\rangle _{123} \\
&=&\frac{1}{2}(I_{a}\otimes U_{1}\otimes U_{2}\otimes U_{3})  \nonumber \\
&&[|\chi ^{+}\rangle _{a12}(I|\varphi \rangle _{3})+|\chi ^{-}\rangle
_{a12}(\sigma _{3}|\varphi \rangle _{3})  \nonumber \\
&&+|\varsigma ^{+}\rangle _{a12}(\sigma _{1}|\varphi \rangle
_{3}|)+|\varsigma ^{-}\rangle _{a12}(\sigma _{3}\sigma _{1}|\varphi \rangle
_{3})] \\
&=&\frac{1}{2}[|\Upsilon ^{+}\rangle _{a12}(U_{3}|\varphi \rangle
_{3})+|\Upsilon ^{-}\rangle _{a12}(U_{3}\sigma _{3}|\varphi \rangle _{3})
\nonumber \\
&&+|\Phi ^{+}\rangle _{a12}(U_{3}\sigma _{1}|\varphi \rangle _{3})+|\Phi
^{-}\rangle _{a12}(U_{3}\sigma _{3}\sigma _{1}|\varphi \rangle _{3})]
\end{eqnarray}

where
\begin{eqnarray}
|\Upsilon ^{+}\rangle _{a12} &=&I_{a}\otimes U_{1}\otimes U_{2}|\chi
^{+}\rangle _{a12}, \\
|\Upsilon ^{-}\rangle _{a12} &=&I_{a}\otimes U_{1}\otimes U_{2}|\chi
^{-}\rangle _{a12}, \\
|\Phi ^{+}\rangle _{a12} &=&I_{a}\otimes U_{1}\otimes U_{2}|\varsigma
^{+}\rangle _{a12}, \\
|\Phi ^{-}\rangle _{a12} &=&I_{a}\otimes U_{1}\otimes U_{2}|\varsigma
^{-}\rangle _{a12}.
\end{eqnarray}

We next show that $|\Upsilon ^{\pm }\rangle _{a12}$ and $|\Phi ^{\pm
}\rangle _{a12}$ are a set of orthogonal states below. A calculation yields

\begin{eqnarray*}
\langle \Upsilon ^{+}|\Upsilon ^{-}\rangle &=&\langle \chi
^{+}|(I_{a}\otimes U_{1}^{\dag }\otimes U_{2}^{\dag })(I_{a}\otimes
U_{1}\otimes U_{2})|\chi ^{-}\rangle \\
&=&\langle \chi ^{+}|\chi ^{-}\rangle =0.
\end{eqnarray*}

Similarly, $|\Phi ^{+}\rangle _{a12}$ and $|\Phi ^{-}\rangle _{a12}$, $%
|\Upsilon ^{+}\rangle _{a12}$ and $|\Phi ^{+}\rangle _{a12}$, $|\Upsilon
^{+}\rangle _{a12}$ and $|\Phi ^{-}\rangle _{a12}$, $|\Upsilon ^{-}\rangle
_{a12}$ and $|\Phi ^{+}\rangle _{a12}$, and $|\Upsilon ^{-}\rangle _{a12}$
and $|\Phi ^{-}\rangle _{a12}$ are orthogonal. Alice can perform a von
Neumann joint measurement on the three qubits `a12'\ by using a set of
orthogonal states\ $|\Upsilon ^{\pm }\rangle _{a12}$ and $|\Phi ^{\pm
}\rangle _{a12}$ and then convey her result to Bob. Then, Bob can convert
his state to the original unknown state\ by applying the following
corresponding unitary transformations: $U_{3}^{\dag }(U_{3}|\varphi \rangle
_{3})=|\varphi \rangle _{3}$, $\sigma _{3}U_{3}^{\dag }(U_{3}\sigma
_{3}|\varphi \rangle _{3})=|\varphi \rangle _{3}$, $\sigma _{1}U_{3}^{\dag
}(U_{3}\sigma _{1}|\varphi \rangle _{3})=|\varphi \rangle _{3}$, and $\sigma
_{1}\sigma _{3}U_{3}^{\dag }(U_{3}\sigma _{3}\sigma _{1}|\varphi \rangle
_{3})=|\varphi \rangle _{3}$. Therefore, $|\psi ^{SD}\rangle _{123}$ is
suitable for\ PTP via the orthogonal states $|\Upsilon ^{\pm }\rangle _{a12}$
and $|\Phi ^{\pm }\rangle _{a12}$.

Note that it is necessary that Bob knows the shared state $|\psi \rangle
_{123}$ in advance. Then, Bob knows the unitary $U_{3}$ which is used to get
$|\psi ^{SD}\rangle _{123}$.

Conversely, assume that $|\psi ^{SD}\rangle _{123}$ is suitable for\ PTP.
Then, by Agrawal-Pati protocol, we can obtain
\begin{eqnarray}
&&|\varphi \rangle _{a}|\psi ^{SD}\rangle _{123} \\
&=&\frac{1}{2}[|\upsilon ^{+}\rangle _{a12}(I|\varphi \rangle
_{3})+|\upsilon ^{-}\rangle _{a12}(\sigma _{3}|\varphi \rangle _{3}) \\
&&+|\nu ^{+}\rangle _{a12}(\sigma _{1}|\varphi \rangle _{3}|)+|\nu
^{-}\rangle _{a12}(\sigma _{3}\sigma _{1}|\varphi \rangle _{3})],
\end{eqnarray}%
where $|\upsilon ^{\pm }\rangle _{a12}$ and $|\nu ^{\pm }\rangle _{a12}$ are
a set of orthogonal states.

Then,

\begin{eqnarray}
&&|\varphi \rangle _{a}|\psi \rangle _{123}  \nonumber \\
&=&I_{a}\otimes U_{1}^{\dag }\otimes U_{2}^{\dag }\otimes U_{3}^{\dag
}|\varphi \rangle _{a}|\psi ^{SD}\rangle _{123} \\
&=&\frac{1}{2}(I_{a}\otimes U_{1}^{\dag }\otimes U_{2}^{\dag }\otimes
U_{3}^{\dag })  \nonumber \\
&&[|\upsilon ^{+}\rangle _{a12}(I|\varphi \rangle _{3})+|\upsilon
^{-}\rangle _{a12}(\sigma _{3}|\varphi \rangle _{3})  \nonumber \\
&&+|\nu ^{+}\rangle _{a12}(\sigma _{1}|\varphi \rangle _{3}|)+|\nu
^{-}\rangle _{a12}(\sigma _{3}\sigma _{1}|\varphi \rangle _{3})] \\
&=&\frac{1}{2}[|\Omega ^{+}\rangle _{a12}(U_{3}^{\dag }|\varphi \rangle
_{3})+|\Omega ^{-}\rangle _{a12}(U_{3}^{\dag }\sigma _{3}|\varphi \rangle
_{3})  \nonumber \\
&&+|\Theta ^{+}\rangle _{a12}(U_{3}^{\dag }\sigma _{1}|\varphi \rangle
_{3})+|\Theta ^{-}\rangle _{a12}(U_{3}^{\dag }\sigma _{3}\sigma _{1}|\varphi
\rangle _{3})],
\end{eqnarray}

where
\begin{eqnarray}
|\Omega ^{+}\rangle _{a12} &=&I_{a}\otimes U_{1}^{\dag }\otimes U_{2}^{\dag
}|\upsilon ^{+}\rangle _{a12}, \\
|\Omega ^{-}\rangle _{a12} &=&I_{a}\otimes U_{1}^{\dag }\otimes U_{2}^{\dag
}|\upsilon ^{-}\rangle _{a12}, \\
|\Theta ^{+}\rangle _{a12} &=&I_{a}\otimes U_{1}^{\dag }\otimes U_{2}^{\dag
}|\nu ^{+}\rangle _{a12}, \\
|\Theta ^{-}\rangle _{a12} &=&I_{a}\otimes U_{1}^{\dag }\otimes U_{2}^{\dag
}|\nu ^{-}\rangle _{a12}.
\end{eqnarray}

Similarly, we can show that $|\Omega ^{\pm }\rangle _{a12}$ and $|\Theta
^{\pm }\rangle _{a12}$ are a set of orthogonal states. Therefore, $|\psi
\rangle _{123}$ is suitable for \ PTP via the orthogonal states $|\Omega
^{\pm }\rangle _{a12}$ and $|\Theta ^{\pm }\rangle _{a12}$. Then, we can
conclude that \ PTP is LU invariant.

\section{Appendix B. For Proposition 3}

\setcounter{equation}{0} \renewcommand{\theequation}{B\arabic{equation}} For
SD in Eq. (\ref{a-bc-sd}), a calculation yields $\rho _{3}=tr_{12}\rho
_{123}=$%
\begin{equation}
\left(
\begin{array}{cc}
\lambda _{1}^{2}+\lambda _{3}^{2} & \lambda _{1}\lambda _{2}e^{i\phi
}+\lambda _{3}\lambda _{4} \\
\lambda _{1}\lambda _{2}e^{-i\phi }+\lambda _{3}\lambda _{4} & \lambda
_{2}^{2}+\lambda _{4}^{2}%
\end{array}%
\right) .
\end{equation}

To make $S(\rho _{3})=\ln 2$, i.e. $\rho _{3}=(1/2)I$, we have

\begin{eqnarray}
\lambda _{1}^{2}+\lambda _{3}^{2} &=&1/2,  \label{a-bc-3} \\
\lambda _{2}^{2}+\lambda _{4}^{2} &=&1/2,  \label{a-bc-4} \\
\lambda _{1}\lambda _{2}e^{i\phi }+\lambda _{3}\lambda _{4} &=&0,
\label{a-bc-5} \\
\lambda _{1}\lambda _{2}e^{-i\phi }+\lambda _{3}\lambda _{4} &=&0.
\label{a-bc-6}
\end{eqnarray}

Case 1. $\lambda _{1}=0$. From Eq. (\ref{a-bc-3}), obtain $\lambda
_{3}^{2}=1/2$. From Eq. (\ref{a-bc-5}), obtain $\lambda _{4}=0$. Then, from
Eq. (\ref{a-bc-4}), obtain $\lambda _{2}^{2}=1/2$. Thus, have the following
state

\begin{equation}
\frac{1}{\sqrt{2}}(|101\rangle +|110\rangle ).  \label{a-bc-7}
\end{equation}

Case 2. $\lambda _{1}\neq 0$. \ Case 2.1. $\lambda _{4}=0$. From Eq. (\ref%
{a-bc-4}), $\lambda _{2}\neq 0$. But, from Eq. (\ref{a-bc-5}), $\lambda
_{2}=0$. It makes a contradiction. So, the subcase cannot happen.

Case 2.2. $\lambda _{4}\neq 0$. We need to consider the following subcases.
Case 2.2.1. $\lambda _{2}=\lambda _{3}=0$. Thus, obtain
\begin{equation}
\frac{1}{\sqrt{2}}(e^{i\phi }|100\rangle +|111\rangle ).  \label{a-bc-8}
\end{equation}

Case 2.2.2. $\lambda _{2}=0$ but $\lambda _{3}\neq 0$. The subcase does not
satisfy Eq. (\ref{a-bc-5}). Therefore, the subcase cannot happen.

Case 2.2.3. $\lambda _{2}\neq 0$ but $\lambda _{3}=0$. From Eq. (\ref{a-bc-5}%
), clearly\ the subcase cannot happen.

Case 2.2.4. $\lambda _{2}\lambda _{3}\neq 0$. For the subcase, $\phi =\pi $
and $\lambda _{1}\lambda _{2}=\lambda _{3}\lambda _{4}$. Via Eqs. (\ref%
{a-bc-3}, \ref{a-bc-4}), obtain $\lambda _{2}=\lambda _{3}$ and $\lambda
_{1}=\lambda _{4}$. Thus, obtain
\begin{equation}
-\lambda _{1}|100\rangle +\lambda _{2}|101\rangle +\lambda _{2}|110\rangle
)+\lambda _{1}|111\rangle ,  \label{a-bc-9}
\end{equation}%
where $\lambda _{1}^{2}+\lambda _{2}^{2}=1/2$ and $\lambda _{1}\lambda
_{2}\neq 0$.

\section{Appendix C. Perfect teleportation for two qubits}

\setcounter{equation}{0} \renewcommand{\theequation}{C\arabic{equation}}

Let $|\psi \rangle _{12}=\sum_{i=0}^{3}c_{i}|i\rangle _{123}$. $|\varphi
\rangle _{a}|\psi \rangle _{12}$ can be rewritten as follows.

\begin{eqnarray*}
&&|\varphi \rangle _{a}|\psi \rangle _{12} \\
&=&\frac{1}{2}[|\xi ^{+}\rangle _{a1}(\alpha |0\rangle _{2}+\beta |1\rangle
_{2})+|\xi ^{-}\rangle _{a1}(\alpha |0\rangle _{2}-\beta |1\rangle _{2}) \\
&&+|\mu ^{+}\rangle _{a1}(\beta |0\rangle _{2}+\alpha |1\rangle _{2})+|\mu
^{-}\rangle _{a1}(\beta |0\rangle _{2}-\alpha |1\rangle _{2})],
\end{eqnarray*}%
where
\begin{eqnarray*}
|\xi ^{\pm }\rangle _{a1} &=&(c_{0}|00\rangle +c_{2}|01\rangle \pm
c_{1}|10\rangle \pm c_{3}|11\rangle )_{a1}, \\
|\mu ^{\pm }\rangle _{a1} &=&(\pm c_{1}|00\rangle \pm c_{3}|01\rangle
+c_{0}|10\rangle +c_{2}|11\rangle )_{a1}.
\end{eqnarray*}%
We can show that the states $|\xi ^{\pm }\rangle _{a1}$ and $|\mu ^{\pm
}\rangle _{a1}$\ are mutually orthogonal if and only if
\begin{eqnarray}
|c_{0}|^{2}+|c_{2}|^{2} &=&|c_{1}|^{2}+|c_{3}|^{2}=\frac{1}{2},
\label{2-qubit-1} \\
c_{0}c_{1}^{\ast }+c_{2}c_{3}^{\ast } &=&0  \label{2-qubit-2}
\end{eqnarray}

Thus, $|\psi \rangle _{12}$ can be used as a resource state for Bennett et
al.'s teleportation if and only if Eqs. (\ref{2-qubit-1}, \ref{2-qubit-2})
hold. In the following subsection of this Appendix, the states satisfying
Eqs. (\ref{2-qubit-1}, \ref{2-qubit-2}) are reduced to the following $|\psi
_{i}\rangle _{12}$, $i=1,2,3$.
\begin{eqnarray}
|\psi _{1}\rangle _{12} &=&\frac{1}{\sqrt{2}}(e^{i\omega _{1}}|01\rangle
+e^{i\omega _{2}}|10\rangle ), \\
|\psi _{2}\rangle _{12} &=&\frac{1}{\sqrt{2}}(e^{i\omega _{0}}|00\rangle
+e^{i\omega _{3}}|11\rangle ), \\
|\psi _{3}\rangle _{12} &=&ke^{i\theta _{0}}|00\rangle +\ell e^{i\theta
_{1}}|01\rangle +\ell e^{i\theta _{2}}|10\rangle +ke^{i\theta
_{3}}|11\rangle ,
\end{eqnarray}%
where $\theta _{0}+\theta _{3}-(\theta _{1}+\theta _{2})=\pm \pi $ and $%
k^{2}+\ell ^{2}=1/2$ and $k\ell \neq 0$.

$|\psi _{i}\rangle _{12}$, $i=1,2,3$, are all the two-qubit states which are
suitable for Bennett et al.'s teleportation. It is easy to check that $|\psi
_{i}\rangle _{12}$, $i=1,2,3$, are LU equivalent to Bell state and have the
maximal concurrence. Thus, Bennett et al.'s teleportation is LU invariant.

From the following subsections, clearly any two of the following three
conditions are equivalent.

(i). The state is one of $|\psi _{i}\rangle _{12}$, $i=1,2,3$, (ii).\ The
state has the maximal concurrence, and (iii). The state is LU equivalent to
Bell state

\subsection{All the two-qubit states satisfying Eqs. (\protect\ref{2-qubit-1}%
, \protect\ref{2-qubit-2})}

Let $|\psi \rangle _{12}=\sum_{i=0}^{3}c_{i}|i\rangle _{123}$. The states
satisfying the Eqs. (\ref{2-qubit-1}, \ref{2-qubit-2}) can be reduced as
follows.

There are three cases.

Case 1. Assume that $c_{0}=0$. Then, $c_{2}\neq 0$ from Eq. (\ref{2-qubit-1}%
). Then, $c_{3}=0$ from Eq. (\ref{2-qubit-2}). Then, $c_{1}\neq 0$ from Eq. (%
\ref{2-qubit-1}). From Eq. (\ref{2-qubit-1}), obtain $%
|c_{1}|^{2}=|c_{2}|^{2}=1/2$. Thus, $|\psi _{1}\rangle _{12}=c_{1}|01\rangle
+c_{2}|10\rangle =\frac{1}{\sqrt{2}}(e^{i\theta _{1}}|01\rangle +e^{i\theta
_{2}}|10\rangle )$.

Case 2. Assume that $c_{2}=0$. Similarly, obtain $|\psi _{2}\rangle
_{12}=c_{0}|00\rangle +c_{3}|11\rangle =\frac{1}{\sqrt{2}}(e^{i\theta
_{0}}|00\rangle +e^{i\theta _{3}}|11\rangle )$.

Case 3. Assume that $c_{i}\neq 0$, $i=0,1,2,3$. Then, from Eqs. (\ref%
{2-qubit-1}, \ref{2-qubit-2}), $|c_{0}|^{2}=|c_{3}|^{2}$ and $%
|c_{2}|^{2}=|c_{1}|^{2}$. Then, obtain $|\psi _{3}\rangle _{12}=ke^{i\theta
_{0}}|00\rangle +\ell e^{i\theta _{1}}|01\rangle +\ell e^{i\theta
_{2}}|10\rangle +ke^{i\theta _{3}}|11\rangle $, where $\theta _{0}+\theta
_{3}-(\theta _{1}+\theta _{2})=\pm \pi $ and $k^{2}+\ell ^{2}=1/2$.

Thus, we find all two-qubit states which are suitable for perfect
teleportation. These states are $|\psi _{i}\rangle _{12}$, $i=1,2,3$. It is
easy to check that $|\psi _{i}\rangle _{12}$, $i=1,2,3$, are LU equivalent
to Bell state and have the maximal concurrence.

\subsection{All the two-qubit states having the maximal concurrence of 1}

Let $c_{i}=r_{i}e^{i\theta _{i}}$, $i=0,1,2,3$. Then, the concurrence of $%
|\psi \rangle _{12}$ is \ that $C(|\psi \rangle
_{12})=2|c_{0}c_{3}-c_{1}c_{2}|$. Clearly,\ Bell state has the maximal
concurrence of 1. We want to explore what states have the maximal
concurrence, i.e. $|c_{0}c_{3}-c_{1}c_{2}|=1/2$, besides Bell state. Let $%
f=|c_{0}c_{3}-c_{1}c_{2}|^{2}$. A calculation yields that $%
f=r_{0}^{2}r_{3}^{2}+r_{1}^{2}r_{2}^{2}-2r_{0}r_{1}r_{2}r_{3}\cos \omega $,
where $\omega =(\theta _{0}+\theta _{3})-(\theta _{1}+\theta _{2})$ and $%
\sum_{i=0}^{3}r_{i}^{2}=1$. One can know that a state has the maximal
concurrence if and only if $f$ has the maximum $1/4$ under the condition $%
\sum_{i=0}^{3}r_{i}^{2}=1$. There are three cases.

Case 1. $r_{0}=0$. Thus, $f=r_{1}^{2}r_{2}^{2}\leq \left( \frac{%
r_{1}^{2}+r_{2}^{2}}{2}\right) ^{2}=\left( \frac{1-r_{3}^{2}}{2}\right)
^{2}\leq 1/4$. When $r_{3}\neq 0$, $f\leq \left( \frac{1-r_{3}^{2}}{2}%
\right) ^{2}<1/4$. When $r_{3}=0$, then $r_{1}^{2}+r_{2}^{2}=1$ and $%
f=r_{1}^{2}r_{2}^{2}\leq \left( \frac{r_{1}^{2}+r_{2}^{2}}{2}\right) ^{2}$ $=%
\frac{1}{4}$ and the equality holds only when $r_{1}=r_{2}$. One can see
that $f=$ $r_{1}^{2}r_{2}^{2}=1/4$ under the condition $%
r_{1}^{2}+r_{2}^{2}=1 $ if and only if $r_{1}=r_{2}=1/\sqrt{2}$. Thus, the
state is $|\psi _{1}\rangle _{12}=\frac{1}{\sqrt{2}}(e^{i\theta
_{1}}|01\rangle +e^{i\theta _{2}}|10\rangle )$.

Case 2. $r_{1}=0$. $f=r_{0}^{2}r_{3}^{2}\leq \left( \frac{r_{0}^{2}+r_{3}^{2}%
}{2}\right) ^{2}=\left( \frac{1-r_{2}^{2}}{2}\right) ^{2}\leq 1/4$.
Similarly, $f=1/4$ implies $r_{2}=0$ and $r_{0}=r_{3}=1/\sqrt{2}$. Thus, the
state is $|\psi _{2}\rangle _{12}=\frac{1}{\sqrt{2}}(e^{i\theta
_{0}}|00\rangle +e^{i\theta _{3}}|11\rangle ).$

Case 3. $r_{i}\neq 0$, $i=0,1,2,3$. Clearly, when $\cos \omega =-1$, i.e. $%
\omega =\pm \pi $, $%
f=r_{0}^{2}r_{3}^{2}+r_{1}^{2}r_{2}^{2}+2r_{0}r_{1}r_{2}r_{3}=(r_{0}r_{3}+r_{1}r_{2})^{2}
$. \ One can see that $r_{0}r_{3}\leq \frac{r_{0}^{2}+r_{3}^{2}}{2}$ and the
equality holds only when $r_{0}=r_{3}$. Similarly, $r_{1}r_{2}\leq \frac{%
r_{1}^{2}+r_{2}^{2}}{2}$ and the equality holds only when $r_{1}=r_{2}$.
Therefore, $(r_{0}r_{3}+r_{1}r_{2})^{2}\leq \left( \frac{%
r_{0}^{2}+r_{3}^{2}+r_{1}^{2}+r_{2}^{2}}{2}\right) ^{2}=\frac{1}{4}$. When $%
r_{0}=r_{3}$ and $r_{1}=r_{2}$, $f=(r_{0}r_{3}+r_{1}r_{2})^{2}=1/4$. The
state is just $|\psi _{3}\rangle _{12}$. When $\cos \omega \neq -1$, $%
f<(r_{0}r_{3}+r_{1}r_{2})^{2}\leq 1/4$.

\section{Appendix D. PSDC-2 is LU invariant}

\setcounter{equation}{0} \renewcommand{\theequation}{D\arabic{equation}} Let
$|\psi \rangle _{123}=\sum_{i=0}^{7}c_{i}|i\rangle _{123}$ be any state of
three qubits. We next prove that $|\psi \rangle _{123}$ is suitable for
PSDC-2, i.e. it satisfies Eqs. (\ref{1-qubit-1}, \ref{1-qubit-2}), if and
only if $|\psi ^{SD}\rangle _{123}$ is also.

Proof. Assume $|\psi \rangle _{123}$ is suitable for PSDC-2. By Theorem 4, $%
|\psi \rangle _{123}$ satisfies Eqs. (\ref{1-qubit-1}, \ref{1-qubit-2}).
Next we show $|\psi ^{SD}\rangle _{123}$ is suitable for PSDC-2. Let $|\psi
^{SD}\rangle _{123}=U_{1}\otimes U_{2}\otimes U_{3}|\psi \rangle _{123}$.
Then, $|\psi \rangle _{123}=U_{1}^{\dag }\otimes U_{2}^{\dag }\otimes
U_{3}^{\dag }|\psi ^{SD}\rangle _{123}$. There are three cases for $U_{1}$,
which is a solution of Eq. (8) in \cite{Dli-jpa-20}. Case 1. $U_{1}=I$. Case
2. $U_{1}=\sigma _{1}$. and Case 3.

\begin{equation}
U_{1}=\left(
\begin{array}{cc}
\frac{t}{m} & \frac{1}{m} \\
\frac{1}{m} & -\frac{t^{\ast }}{m}%
\end{array}%
\right) ,  \label{U-A-2}
\end{equation}%
where $m=\sqrt{|t|^{2}+1}$.

Let

\begin{eqnarray}
A_{1}|\psi ^{SD}\rangle _{123} &=&I\otimes I\otimes I|\psi ^{SD}\rangle
_{123}=|\alpha ^{+}\rangle , \\
A_{2}|\psi ^{SD}\rangle _{123} &=&\sigma _{3}\otimes I\otimes I|\psi
^{SD}\rangle _{123}=|\alpha ^{-}\rangle , \\
A_{3}|\psi ^{SD}\rangle _{123} &=&\sigma _{1}\otimes I\otimes I|\psi
^{SD}\rangle _{123}=|\beta ^{+}\rangle , \\
A_{4}|\psi ^{SD}\rangle _{123} &=&\sigma _{3}\sigma _{1}\otimes I\otimes
I|\psi ^{SD}\rangle _{123}=|\beta ^{-}\rangle .
\end{eqnarray}%
Then, we need to show that the states $|\alpha ^{+}\rangle $, $|\alpha
^{-}\rangle $, $|\beta ^{+}\rangle $, and $|\beta ^{-}\rangle $ are mutually
orthogonal by computing six dot products of the states $|\alpha ^{+}\rangle $%
, $|\alpha ^{-}\rangle $, $|\beta ^{+}\rangle $, and $|\beta ^{-}\rangle $.
A calculation yields that $\langle \alpha ^{-}|\beta ^{+}\rangle =\langle
\alpha ^{+}|\beta ^{-}\rangle $, $\langle \alpha ^{-}|\beta ^{-}\rangle
=\langle \alpha ^{+}|\beta ^{+}\rangle $, and $\langle \beta ^{+}|\beta
^{-}\rangle =-\langle \alpha ^{+}|\alpha ^{-}\rangle $. Therefore, we only
need to show that $\langle \alpha ^{+}|\beta ^{-}\rangle $, $\langle \alpha
^{+}|\beta ^{+}\rangle $, and $\langle \alpha ^{+}|\alpha ^{-}\rangle $
vanish.

Step 1. Let us compute
\begin{equation}
\langle \alpha ^{+}|\alpha ^{-}\rangle =\langle \psi ^{SD}|(\sigma
_{3}\otimes I\otimes I)|\psi ^{SD}\rangle _{123}=\langle \psi |(U_{1}^{\dag
}\sigma _{3}U_{1}\otimes I\otimes I)|\psi \rangle .  \label{step-1}
\end{equation}

Case 1.1. $U_{1}=I$. Then, $U_{1}^{\dag }\sigma _{3}U_{1}=\sigma _{3}$.
Then, Eq. (\ref{step-1}) reduces to
\begin{equation}
\langle \alpha ^{+}|\alpha ^{-}\rangle =\langle \psi |(\sigma _{3}\otimes
I\otimes I)|\psi \rangle  \label{step-1-1}
\end{equation}

A calculation yields that $(\sigma _{3}\otimes I\otimes I)|\psi \rangle =$

$c_{0}|000\rangle +c_{1}|001\rangle +c_{2}|010\rangle +c_{3}|011\rangle
-c_{4}|100\rangle -c_{5}|101\rangle -c_{6}|110\rangle -c_{7}|111\rangle $.

Then, from Eq. (\ref{step-1-1}), $\langle \alpha ^{+}|\alpha ^{-}\rangle
=\sum_{i=0}^{3}|c_{i}|^{2}-\sum_{i=4}^{7}|c_{i}|^{2}$. Since $|\psi \rangle $
is suitable for PSDC-2, by Eq. (\ref{1-qubit-1}) $\langle \alpha ^{+}|\alpha
^{-}\rangle =0$.

Case 1.2. $U_{1}=\sigma _{1}$. Then, $U_{1}^{\dag }\sigma _{3}U_{1}=\sigma
_{1}\sigma _{3}\sigma _{1}=-\sigma _{3}$. Thus, from Eq. (\ref{step-1}), $%
\langle \alpha ^{+}|\alpha ^{-}\rangle =-\langle \psi |(\sigma _{3}\otimes
I\otimes I)|\psi \rangle =0$ by Case 1.1.

Case 1.3. Consider $U_{1}$ in Eq. (\ref{U-A-2}). Then,%
\begin{equation}
U_{1}^{\dag }\sigma _{3}U_{1}=\allowbreak \left(
\begin{array}{cc}
\frac{1}{m^{2}}tt^{\ast }-\frac{1}{m^{2}} & \frac{2}{m^{2}}t^{\ast } \\
\frac{2}{m^{2}}t & \frac{1}{m^{2}}-\frac{1}{m^{2}}tt^{\ast }%
\end{array}%
\right) .
\end{equation}

Then,

\begin{eqnarray}
U_{1}^{\dag }\sigma _{3}U_{1}|0\rangle &=&(\frac{1}{m^{2}}tt^{\ast }-\frac{1%
}{m^{2}})|0\rangle +\frac{2}{m^{2}}t|1\rangle  \label{step-1-3-1} \\
U_{1}^{\dag }\sigma _{3}U_{1}|1\rangle &=&\frac{2}{m^{2}}t^{\ast }|0\rangle
+(\frac{1}{m^{2}}-\frac{1}{m^{2}}tt^{\ast })|1\rangle .  \label{step-1-3-2}
\end{eqnarray}

Via Eqs. (\ref{step-1-3-1}, \ref{step-1-3-2}), from Eq. (\ref{step-1}) a
complicated calculation yields that $\langle \alpha ^{+}|\alpha ^{-}\rangle
=0$ by Theorem 4.

Step 2. Let us compute
\begin{equation}
\langle \alpha ^{+}|\beta ^{+}\rangle =\langle \psi ^{SD}|(\sigma
_{1}\otimes I\otimes I)|\psi ^{SD}\rangle _{123}=\langle \psi |(U_{1}^{\dag
}\sigma _{1}U_{1}\otimes I\otimes I)|\psi \rangle .  \label{step-2}
\end{equation}

Case 2.1. $U_{1}=I$. Then, $U_{1}^{\dag }\sigma _{1}U_{1}=\sigma _{1}$.
Then, Eq. (\ref{step-2}) reduces to
\begin{equation}
\langle \alpha ^{+}|\beta ^{+}\rangle =\langle \psi |(\sigma _{1}\otimes
I\otimes I)|\psi \rangle .  \label{step-2-1}
\end{equation}

A calculation yields that $(\sigma _{1}\otimes I\otimes I)|\psi \rangle =$

$c_{0}|100\rangle +c_{1}|101\rangle +c_{2}|110\rangle +c_{3}|111\rangle
+c_{4}|000\rangle +c_{5}|001\rangle +c_{6}|010\rangle +c_{7}|011\rangle $.

Then, from Eq. (\ref{step-2-1}) $\langle \alpha ^{+}|\beta ^{+}\rangle =$

$c_{0}^{\ast }c_{4}+c_{1}^{\ast }c_{5}+c_{2}^{\ast }c_{6}+c_{3}^{\ast
}c_{7}+c_{4}^{\ast }c_{0}+c_{5}^{\ast }c_{1}+c_{6}^{\ast }c_{2}+c_{7}^{\ast
}c_{3}=0$ by Eq. (\ref{1-qubit-2}).

Case 2.2. $U_{1}=\sigma _{1}$. Then, $U_{1}^{\dag }\sigma _{1}U_{1}=\sigma
_{1}\sigma _{1}\sigma _{1}=\sigma _{1}$. By Case 2.1, $\langle \alpha
^{+}|\beta ^{+}\rangle =0$.

Case 2.3. Consider $U_{1}$ in Eq. (\ref{U-A-2}). Then,

\begin{equation}
U_{1}^{\dag }\sigma _{1}U_{1}\allowbreak =\allowbreak \left(
\begin{array}{cc}
\frac{1}{m^{2}}\left( t+t^{\ast }\right) & -\frac{1}{m^{2}}\left( (t^{\ast
})^{2}-1\right) \\
-\frac{1}{m^{2}}\left( t^{2}-1\right) & -\frac{1}{m^{2}}\left( t+t^{\ast
}\right)%
\end{array}%
\right) .
\end{equation}

Then,

\begin{eqnarray}
U_{1}^{\dag }\sigma _{1}U_{1}|0\rangle &=&\frac{1}{m^{2}}\left( t+t^{\ast
}\right) |0\rangle -\frac{1}{m^{2}}\left( t^{2}-1\right) |1\rangle ,
\label{step-2-3-1} \\
U_{1}^{\dag }\sigma _{1}U_{1}|1\rangle &=&-\frac{1}{m^{2}}\left( (t^{\ast
})^{2}-1\right) |0\rangle -\frac{1}{m^{2}}\left( t+t^{\ast }\right)
|1\rangle .  \label{step-2-3-2}
\end{eqnarray}

Via Eqs. (\ref{step-2-3-1}, \ref{step-2-3-2}), from Eq. (\ref{step-2}) a
complicated calculation yields that $\langle \alpha ^{+}|\beta ^{+}\rangle
=0 $ by Theorem 4.

Step 3. Let us compute

\begin{equation}
\langle \alpha ^{+}|\beta ^{-}\rangle =\langle \psi ^{SD}|\sigma _{3}\sigma
_{1}\otimes I\otimes I|\psi ^{SD}\rangle =\langle \psi |(U_{1}^{\dag }\sigma
_{3}\sigma _{1}U_{1}\otimes I\otimes I)|\psi \rangle .  \label{step-3}
\end{equation}

Case 3.1. $U_{1}=I$. Then, $U_{1}^{\dag }\sigma _{3}\sigma _{1}U_{1}=\sigma
_{3}\sigma _{1}$. Then, Eq. (\ref{step-3}) reduces to
\begin{equation}
\langle \alpha ^{+}|\beta ^{-}\rangle =\langle \psi |(\sigma _{3}\sigma
_{1}\otimes I\otimes I)|\psi \rangle .  \label{step-3-1}
\end{equation}

A calculation yields $(\sigma _{3}\sigma _{1}\otimes I\otimes I)|\psi
\rangle =$

$-c_{0}|100\rangle -c_{1}|101\rangle -c_{2}|110\rangle -c_{3}|111\rangle
+c_{4}|000\rangle +c_{5}|001\rangle +c_{6}|1010\rangle +c_{7}|011\rangle $.

From Eq. (\ref{step-3-1}), a calculation yields that $\langle \alpha
^{+}|\beta ^{-}\rangle =c_{0}^{\ast }c_{4}+c_{1}^{\ast }c_{5}+c_{2}^{\ast
}c_{6}+c_{3}^{\ast }c_{7}-c_{0}c_{4}^{\ast }-c_{1}c_{5}^{\ast
}-c_{2}c_{6}^{\ast }-c_{3}c_{7}^{\ast }=0$ by Theorem 4.

Case 3.2. $U_{1}=\sigma _{1}$. Then, $U_{1}^{\dag }\sigma _{3}\sigma
_{1}U_{1}=\sigma _{1}\sigma _{3}\sigma _{1}\sigma _{1}=\sigma _{1}\sigma
_{3}=-\sigma _{3}\sigma _{1}$. Ref. Case 3.1., $\langle \alpha ^{+}|\beta
^{-}\rangle =0.$

Case 3.3. Consider $U_{1}$ in Eq. (\ref{U-A-2}). Then,

\begin{equation}
U_{1}^{\dag }\sigma _{3}\sigma _{1}U_{1}=\allowbreak \left(
\begin{array}{cc}
\frac{1}{m^{2}}t^{\ast }-\frac{1}{m^{2}}t & -\frac{1}{m^{2}}(t^{\ast })^{2}-%
\frac{1}{m^{2}} \\
\frac{1}{m^{2}}t^{2}+\frac{1}{m^{2}} & \frac{1}{m^{2}}t-\frac{1}{m^{2}}%
t^{\ast }%
\end{array}%
\right)
\end{equation}%
Then,

\begin{eqnarray}
U_{1}^{\dag }\sigma _{3}\sigma _{1}U_{1}|0\rangle &=&(\frac{1}{m^{2}}t^{\ast
}-\frac{1}{m^{2}}t)|0\rangle +(\frac{1}{m^{2}}t^{2}+\frac{1}{m^{2}}%
)|1\rangle ,  \label{step-3-3-1} \\
U_{1}^{\dag }\sigma _{3}\sigma _{1}U_{1}|1\rangle &=&(-\frac{1}{m^{2}}%
(t^{\ast })^{2}-\frac{1}{m^{2}})|0\rangle +(\frac{1}{m^{2}}t-\frac{1}{m^{2}}%
t^{\ast })|1\rangle .  \label{step-3-3-2}
\end{eqnarray}

Via Eqs. (\ref{step-3-3-1}, \ref{step-3-3-2}), from Eq. (\ref{step-3}) a
complicated calculation yields that $\langle \alpha ^{+}|\beta ^{-}\rangle
=0 $ by Theorem 4.

Conversely, clearly it is also true that if $|\psi ^{SD}\rangle $ is
suitable for PSDC-2, then $|\psi \rangle $ is also.

\section{Appendix E Superdense coding}

\setcounter{equation}{0} \renewcommand{\theequation}{E\arabic{equation}}

\subsection{For PSDC-2}

Let $I\otimes I\otimes I\sum_{i=0}^{7}c_{i}|i\rangle _{123}=|T^{+}\rangle $.
Then, a calculation yields $|T^{+}\rangle =$

$c_{0}|000\rangle +c_{1}|001\rangle +c_{2}|010\rangle +c_{3}|011\rangle
+c_{4}|100\rangle +c_{5}|101\rangle +c_{6}|110\rangle +c_{7}|111\rangle $.

Let $\sigma _{3}\otimes I\otimes I\sum_{i=0}^{7}c_{i}|i\rangle
_{123}=|T^{-}\rangle $. Then, $|T^{-}\rangle =$

$c_{0}|000\rangle +c_{1}|001\rangle +c_{2}|010\rangle +c_{3}|011\rangle
-c_{4}|100\rangle -c_{5}|101\rangle -c_{6}|110\rangle -c_{7}|111\rangle $.

Let $\sigma _{1}\otimes I\otimes I\sum_{i=0}^{7}c_{i}|i\rangle
_{123}=|H^{+}\rangle $. Then, $|H^{+}\rangle =$

$c_{0}|100\rangle +c_{1}|101\rangle +c_{2}|110\rangle +c_{3}|111\rangle
+c_{4}|000\rangle +c_{5}|001\rangle +c_{6}|010\rangle +c_{7}|011\rangle $.

Let $\sigma _{3}\sigma _{1}\otimes I\otimes I\sum_{i=0}^{7}c_{i}|i\rangle
_{123}=|H^{-}\rangle $. Then, $|H^{-}\rangle =$

$-c_{0}|100\rangle -c_{1}|101\rangle -c_{2}|110\rangle -c_{3}|111\rangle
+c_{4}|000\rangle +c_{5}|001\rangle +c_{6}|010\rangle +c_{7}|011\rangle $.

A calculation shows that $\langle H^{+}|H^{-}\rangle =-\langle
T^{+}|T^{-}\rangle $, $\langle T^{-}|H^{-}\rangle =\langle
T^{+}|H^{+}\rangle $, $\langle T^{-}|H^{+}\rangle =\langle
T^{+}|H^{-}\rangle $. So, we only need \ to compute three dot products below.

From that $\sum_{i=0}^{7}|c_{i}|^{2}=1$, then $\langle T^{+}|T^{-}\rangle =0$
if and only if
\begin{eqnarray}
\sum_{i=0}^{3}|c_{i}|^{2} &=&1/2,  \label{sdc-2-1} \\
\sum_{i=4}^{7}|c_{i}|^{2} &=&1/2.  \label{sdc-2-2}
\end{eqnarray}

One can argue that $\langle T^{+}|H^{+}\rangle =0$ and $\langle
T^{+}|H^{-}\rangle =0$ if and only if
\begin{equation}
c_{0}^{\ast }c_{4}+c_{1}^{\ast }c_{5}+c_{2}^{\ast }c_{6}+c_{3}^{\ast
}c_{7}=0.  \label{sdc-2-3}
\end{equation}

\subsection{For PSDC-3}

Let $I\otimes \sigma _{1}\otimes I\sum_{i=0}^{7}c_{i}|i\rangle
_{123}=|L^{+}\rangle $. Then, $|L^{+}\rangle =$

$c_{0}|010\rangle +c_{1}|011\rangle +c_{2}|000\rangle +c_{3}|001\rangle
+c_{4}|110\rangle +c_{5}|111\rangle +c_{6}|100\rangle +c_{7}|101\rangle $.

Let $I\otimes \sigma _{3}\sigma _{1}\otimes I\sum_{i=0}^{7}c_{i}|i\rangle
_{123}=|L^{-}\rangle $. Then, $|L^{-}\rangle =$

$-c_{0}|010\rangle -c_{1}|011\rangle +c_{2}|000\rangle +c_{3}|001\rangle
-c_{4}|110\rangle -c_{5}|111\rangle +c_{6}|100\rangle +c_{7}|101\rangle $.

Let $\sigma _{1}\otimes \sigma _{1}\otimes I\sum_{i=0}^{7}c_{i}|i\rangle
_{123}=|D^{+}\rangle $. Then, $|D^{+}\rangle =$

$c_{0}|110\rangle +c_{1}|111\rangle +c_{2}|100\rangle +c_{3}|101\rangle
+c_{4}|010\rangle +c_{5}|011\rangle +c_{6}|000\rangle +c_{7}|001\rangle $.

Let $\sigma _{1}\otimes \sigma _{3}\sigma _{1}\otimes
I\sum_{i=0}^{7}c_{i}|i\rangle _{123}=|D^{-}\rangle $. Then, $|D^{-}\rangle =$

$-c_{0}|110\rangle -c_{1}|111\rangle +c_{2}|100\rangle +c_{3}|101\rangle
-c_{4}|010\rangle -c_{5}|011\rangle +c_{6}|000\rangle +c_{7}|001\rangle $.

\subsubsection{$|L^{\pm }\rangle $ and $|D^{\pm }\rangle $ are a set of
orthogonal states}

We can show that $\langle L^{+}|L^{-}\rangle =\langle D^{+}|D^{-}\rangle $, $%
\langle L^{-}|D^{+}\rangle =\langle L^{+}|D^{-}\rangle $, and $\langle
L^{-}|D^{-}\rangle =\langle L^{+}|D^{+}\rangle $. Thus, we only need to
compute the following three dot products.

From that $\sum_{i=0}^{7}|c_{i}|^{2}=1$, then $\langle L^{+}|L^{-}\rangle =0$
if and only if
\begin{eqnarray}
|c_{2}|^{2}+|c_{3}|^{2}+|c_{6}|^{2}+|c_{7}|^{2} &=&1/2,  \label{coe-a} \\
|c_{0}|^{2}+|c_{1}|^{2}+|c_{4}|^{2}+|c_{5}|^{2} &=&1/2.  \label{coe-b}
\end{eqnarray}

From Eqs. (\ref{sdc-2-1}, \ref{coe-a}) and Eqs. (\ref{sdc-2-2}, \ref{coe-b}%
), obtain, respectively
\begin{eqnarray}
|c_{0}|^{2}+|c_{1}|^{2} &=&|c_{6}|^{2}+|c_{7}|^{2},  \label{coe-c} \\
|c_{2}|^{2}+|c_{3}|^{2} &=&|c_{4}|^{2}+|c_{5}|^{2}.  \label{coe-d}
\end{eqnarray}

One can see that $\langle L^{+}|D^{-}\rangle =0$ and $\langle
L^{+}|D^{+}\rangle =0$ if and only if
\begin{eqnarray}
\mathrm{Re} (c_{2}^{\ast }c_{6}+c_{3}^{\ast }c_{7}) &=&0,  \label{coe-e} \\
\mathrm{Re} (c_{0}^{\ast }c_{4}+c_{1}^{\ast }c_{5}) &=&0.  \label{coe-f}
\end{eqnarray}

Clearly, Eqs. (\ref{sdc-2-3}, \ref{coe-e}, \ref{coe-f}) are equivalent to
Eqs. (\ref{sdc-2-3}, \ref{coe-e}).

\subsubsection{$|L^{\pm }\rangle $ and $|D^{\pm }\rangle $ are orthogonal to
$|T^{\pm }\rangle $ and $|H^{\pm }\rangle $}

We need to make that $|L^{\pm }\rangle $ and $|D^{\pm }\rangle $ are
orthogonal to $|T^{\pm }\rangle $ and $|H^{\pm }\rangle $ below. A
calculation yields the following equalities.

$\langle D^{+}|T^{+}\rangle =\langle L^{+}|H^{+}\rangle $, $\langle
D^{+}|T^{-}\rangle =-\langle L^{+}|H^{-}\rangle $, $\langle
D^{+}|H^{+}\rangle =\langle L^{+}|T^{+}\rangle $,

$\langle D^{+}|H^{-}\rangle =-\langle L^{+}|T^{-}\rangle $, $\langle
D^{-}|T^{+}\rangle =\langle L^{-}|H^{+}\rangle $, $\langle
D^{-}|T^{-}\rangle =-\langle L^{-}|H^{-}\rangle $,

$\langle D^{-}|H^{+}\rangle =\langle L^{-}|T^{+}\rangle $, $\langle
D^{-}|H^{-}\rangle =-\langle L^{-}|T^{-}\rangle $.

One can show that $\langle L^{+}|T^{+}\rangle =$ $\langle L^{+}|T^{-}\rangle
=$ $\langle L^{-}|T^{-}\rangle =$ $\langle L^{-}|T^{+}\rangle =0$\ if and
only if
\begin{eqnarray}
c_{0}^{\ast }c_{2}+c_{1}^{\ast }c_{3} &=&0,  \label{uu1} \\
c_{4}^{\ast }c_{6}+c_{5}^{\ast }c_{7} &=&0.  \label{uu2}
\end{eqnarray}

Similarly,a calculation shows that

$\langle L^{+}|H^{+}\rangle =\langle L^{+}|H^{-}\rangle =\langle
L^{-}|H^{+}\rangle =$ $\langle L^{-}|H^{-}\rangle =0$\ if and only if
\begin{eqnarray}
c_{2}c_{4}^{\ast }+c_{3}c_{5}^{\ast } &=&0,  \label{tt1} \\
c_{0}^{\ast }c_{6}+c_{1}^{\ast }c_{7} &=&0.  \label{tt2}
\end{eqnarray}

\section{Appendix F. Solving equations (\protect\ref{x3}-\protect\ref{t2})
for PSDC-3}

\setcounter{equation}{0} \renewcommand{\theequation}{F\arabic{equation}}

Let $|\psi \rangle _{123}=\sum_{i=0}^{7}c_{i}|i\rangle _{123}$ be any state
of three qubits. For $|\psi \rangle _{123}$, 3-tangle $\tau _{123}$ has the
following three versions, which are the same \cite{pla-li}.%
\begin{eqnarray}
&&\tau _{123}  \nonumber \\
&=&4|(c_{0}c_{7}-c_{1}c_{6}+c_{2}c_{5}-c_{3}c_{4})^{2}-4(c_{1}c_{4}-c_{0}c_{5})(c_{3}c_{6}-c_{2}c_{7})|
\\
&=&4|(c_{0}c_{7}+c_{1}c_{6}-c_{2}c_{5}-c_{3}c_{4})^{2}-4(c_{2}c_{4}-c_{0}c_{6})(c_{3}c_{5}-c_{1}c_{7})|
\\
&=&4|(c_{0}c_{7}-c_{1}c_{6}-c_{2}c_{5}+c_{3}c_{4})^{2}-4(c_{0}c_{3}-c_{1}c_{2})(c_{4}c_{7}-c_{5}c_{6})|.
\end{eqnarray}%
\newline

To solve Eqs. (\ref{x3}-\ref{t2}) for PSDC-3, there are the following cases.
(A). $c_{i}\neq 0$, $i=0,...,7$; (B). $c_{0}c_{5}c_{1}c_{4}\neq 0$; (C). $%
c_{0}c_{5}c_{1}c_{4}=0$. There are the following subcases for (C). (C.1).
Only one of $c_{0},c_{5},c_{1},c_{4}$ vanishes, for the case we show that
Eqs. (\ref{x3}-\ref{t2}) for PSDC-3 have no solution; (C.2). Only one of $%
c_{0},c_{5},c_{1},c_{4}$ does not vanish, for the case we show that the
states $|F_{0}\rangle ,|F_{1}\rangle ,|F_{2}\rangle ,$ and $|F_{3}\rangle $
are suitable for PSDC-3; (C.3.) Only two of $c_{0},c_{1},c_{5},$ and $c_{4}$
vanish, for the case we show that $|\pi _{1}\rangle ,|\pi _{2}\rangle ,|\pi
_{3}\rangle ,$ and $|\pi _{4}\rangle $ are suitable for PSDC-3; (C.4). All
of $c_{0},c_{1},c_{5},c_{4}$ vanish, the case cannot happen.

There are six subcases for (C.3). (C.3.1). $c_{0}=c_{1}=0$ \& $%
c_{4}c_{5}\neq 0$; (C.3.2). $c_{4}=c_{5}=0$ \& $c_{0}c_{1}\neq 0$; (C.3.3.).
$c_{0}=c_{5}=0$ \& $c_{1}c_{4}\neq 0$; (C.3.4). $c_{0}=c_{4}=0$ \& $%
c_{1}c_{5}\neq 0$; (C.3.5). $c_{1}=c_{5}=0$ \& $c_{0}c_{4}\neq 0$; (C.3.6). $%
c_{1}=c_{4}=0$ \& $c_{0}c_{5}\neq 0$.

(A). $c_{i}\neq 0$, $i=0,...,7$.

From Eqs. (\ref{u1}-\ref{t2}), obtain

\begin{eqnarray}
c_{0}c_{5} &=&c_{1}c_{4},  \label{wc-5} \\
c_{2}c_{7} &=&c_{3}c_{6}.  \label{wc-6}
\end{eqnarray}

(A.1). The SLOCC\ classes A-BC, B-AC, C-AB, and A-B-C are not suitable for
PSDC-3.

We can show that the SLOCC\ class A-BC (resp. B-AC, C-AB, A-B-C) does not
satisfy Eq. (\ref{a3}) (resp. Eq. (\ref{u1}), Eq. (\ref{u1}), Eq. (\ref{t2}%
)), respectively. So, these SLOCC\ classes are not suitable for PSDC-3.

For example, we prove that any state of the SLOCC\ class C-AB does not
satisfy Eq. (\ref{u1}) as follows.

Let $|\Sigma \rangle _{123}=(a|00\rangle +b|01\rangle +c|10\rangle
+d|11\rangle )_{12}(e|0\rangle +f|1\rangle )_{3}$, where $abcdef\neq 0$
because $c_{i}\neq 0$, $i=0,...,7$. We can rewrite

$|\Sigma \rangle _{123}=ae|000\rangle +af|001\rangle +be|010\rangle
+bf|011\rangle +ce|100\rangle +cf|101\rangle +de|110\rangle +df|111\rangle $.

Then,\ $c_{0}^{\ast }c_{2}+c_{1}^{\ast }c_{3}=a^{\ast }e^{\ast }be+a^{\ast
}f^{\ast }bf=a^{\ast }b(|e|^{2}+|f|^{2})\neq 0$. Therefore, C-AB does not
satisfy Eq. (\ref{u1}).

(A.2). From the table \cite{pla-li}, one can see that any state of W SLOCC\
class satisfies $c_{0}c_{5}\neq c_{1}c_{4}$ or $c_{2}c_{7}\neq c_{3}c_{6}$.
Therefore, any state of W SLOCC\ class does not satisfy Eqs. (\ref{wc-5}, %
\ref{wc-6}). Thus, when $c_{i}\neq 0$, $i=0,...,7$, any state of W SLOCC\
class is not suitable for PSDC-3.

(A.3). We next compute 3-tangle for $|\psi \rangle _{123}$ \cite{pla-li}
\begin{equation}
\tau
_{123}=4|(c_{0}c_{7}-c_{1}c_{6}+c_{2}c_{5}-c_{3}c_{4})^{2}-4(c_{1}c_{4}-c_{0}c_{5})(c_{3}c_{6}-c_{2}c_{7})|.
\end{equation}

From Eqs. (\ref{wc-5}, \ref{wc-6}), obtain
\begin{equation}
\tau _{123}=4|(c_{0}c_{7}-c_{1}c_{6}+c_{2}c_{5}-c_{3}c_{4})^{2}|.
\end{equation}

From Eqs. (\ref{wc-5}, \ref{wc-6}), let
\begin{equation}
\frac{c_{0}}{c_{1}}=\frac{c_{4}}{c_{5}}=p,\frac{c_{2}}{c_{3}}=\frac{c_{6}}{%
c_{7}}=q.  \label{cd-1}
\end{equation}

Via Eq. (\ref{cd-1}), obtain

\begin{equation}
c_{0}c_{7}-c_{1}c_{6}+c_{2}c_{5}-c_{3}c_{4}=(p-q)(c_{1}c_{7}-c_{3}c_{5}).
\end{equation}

When $\tau _{123}=0$, i.e. $\frac{c_{0}}{c_{1}}=\frac{c_{4}}{c_{5}}=\frac{%
c_{2}}{c_{3}}=\frac{c_{6}}{c_{7}}$ or $c_{1}c_{7}=c_{3}c_{5}$, these states
belong to the SLOCC\ classes W, A-BC, B-AC, C-AB, or A-B-C \cite{Dur, pla-li}%
. We show that these states are not suitable for PSDC-3 in (A1) and (A2).
That is, Eqs. (\ref{x3}-\ref{t2}) for PSDC-3 do not have solutions when $%
\tau _{123}=0$.

When $\tau _{123}\neq 0$, i.e.
\begin{eqnarray}
\frac{c_{0}}{c_{1}} &=&\frac{c_{4}}{c_{5}}\neq \frac{c_{2}}{c_{3}}=\frac{%
c_{6}}{c_{7}},  \label{ghz-x} \\
\frac{c_{1}}{c_{3}} &\neq &\frac{c_{5}}{c_{7}},  \label{ghz-y}
\end{eqnarray}%
these states belong to GHZ SLOCC\ class \cite{Dur, pla-li}. Under the
conditions in Eqs. (\ref{ghz-x}, \ref{ghz-y}), we do not know whether or not
Eqs. (\ref{x3}-\ref{t2}) for PSDC-3 have solutions. Anyway, we can say if
Eqs. (\ref{x3}-\ref{t2}) for PSDC-3 have solutions under Eqs. (\ref{ghz-x}, %
\ref{ghz-y}), the solutions must be states of GHZ SLOCC\ class.

(B). $c_{0}c_{5}c_{1}c_{4}\neq 0$.

From Eq. (\ref{u2}), there are two subcases. Case B.1. $c_{6}=c_{7}=0$. It
does not satisfy Eq. (\ref{x3}).

Case B.2. $c_{6}c_{7}\neq 0$. From Eq. (\ref{t1}) and $c_{5}c_{4}\neq 0$,
there are subcases. Case B.2.1. $c_{2}=$ $c_{3}=0$. It does not satisfy Eq. (%
\ref{x4}). Case B.2.2. $c_{2}c_{3}\neq 0$. Thus, $c_{i}\neq 0$, $i=0,...,7$.
Then, it follows (A).

(C). $c_{0}c_{5}c_{1}c_{4}=0$

In the subsection, we show that any state of the SLOCC\ classes W, A-BC,
B-AC, C-AB, and A-B-C is not suitable for PSDC-3, and find all states of GHZ
SLOCC class which are suitable for PSDC-3.

(C.1). Only one of $c_{0},c_{5},c_{1},c_{4}$ vanishes.

We show that any state satisfying the condition is not suitable for PSDC-3.

For example, $c_{0}=0$ \& $c_{5}c_{1}c_{4}\neq 0$. From Eqs. (\ref{u1}, \ref%
{t2}), obtain $c_{3}=c_{7}=0$. From Eqs. (\ref{u2}, \ref{t1}), obtain $%
c_{2}=c_{6}=0$. It does not satisfy Eq. (\ref{x4}).

We omit other cases.

(C.2). Only one of $c_{0},c_{5},c_{1},c_{4}$ does not vanish.

(C.2.1). $c_{0}\neq 0$ \& $c_{5}=c_{1}=c_{4}=0$.

From Eq. (\ref{x4}), obtain $c_{2}=c_{3}=0$. From Eq. (\ref{t2}), $c_{6}=0$.
From Eq. (\ref{x3}), $|c_{0}|=|c_{7}|=1/\sqrt{2}$.

Thus, obtain the state $|F_{0}\rangle =(1/\sqrt{2})(e^{i\theta
_{0}}|000\rangle +e^{i\theta _{7}}|111\rangle )$, which is suitable for
PSDC-3.

(C.2.2). $c_{5}\neq 0$ \& $c_{0}=c_{1}=c_{4}=0$. From Eq. (\ref{x3}), $%
c_{6}=c_{7}=0$. From Eq. (\ref{t1}), $c_{3}=0$. \ Thus, from Eq. (\ref{x4}),
$|c_{2}|=|c_{5}|=1/\sqrt{2}$. Thus, obtain the state $|F_{2}\rangle =(1/%
\sqrt{2})(e^{i\theta _{2e}}|010\rangle +e^{i\theta _{5}}|101\rangle ),$,
which is suitable for PSDC-3.

(C.2.3). $c_{1}\neq 0$ \& $c_{0}=c_{5}=c_{4}=0$.

From Eq. (\ref{x4}), obtain $c_{2}=c_{3}=0$. From Eq. (\ref{t2}), $c_{7}=0$.
From Eq. (\ref{x3}), $|c_{1}|=|c_{6}|=1/\sqrt{2}$. \ Thus, obtain the state $%
|F_{1}\rangle =(1/\sqrt{2})(e^{i\theta _{1}}|001\rangle +e^{i\theta
_{6}}|110\rangle ),$, which is suitable for PSDC-3.

(C.2.4). $c_{4}\neq 0$ \& $c_{0}=c_{5}=c_{1}=0$. From Eq. (\ref{x3}), $%
c_{6}=c_{7}=0$. From Eq. (\ref{t1}), $c_{2}=0$. From Eq. (\ref{x4}), obtain $%
|c_{3}|=|c_{4}|=1/\sqrt{2}$. Thus, obtain the state $|F_{3}\rangle =(1/\sqrt{%
2})(e^{i\theta _{3}}|011\rangle +e^{i\theta _{4}}|100\rangle )$, which is
suitable for PSDC-3.

(C.3.) Only two of $c_{0},c_{1},c_{5},$ and $c_{4}$ vanish.

There are six subcases. $c_{0}=c_{1}=0$, $c_{5}=c_{4}=0,$ $%
c_{0}=c_{5}=0,c_{0}=c_{4}=0,c_{1}=c_{5}=0,c_{1}=c_{4}=0.$

(C.3.1). $c_{0}=c_{1}=0$ \& $c_{4}c_{5}\neq 0$.

If $c_{0}=c_{1}=0$ then $c_{6}=c_{7}=0$ from Eq. (\ref{x3}). From Eq. (\ref%
{t1}), there are two subcases: $c_{2}c_{3}\neq 0$ and $c_{2}c_{3}=0$.

(C.3.1.1). $c_{2}c_{3}\neq 0$. Thus, the state is reduced to $|\pi
_{1}\rangle =c_{2}|010\rangle +c_{3}|011\rangle +c_{4}|100\rangle
+c_{5}|101\rangle $. Then, from Eq. (\ref{t1}), obtain $\frac{c_{2}}{c_{3}}=-%
\frac{c_{5}^{\ast }}{c_{4}^{\ast }}$. Let $\frac{c_{2}}{c_{3}}=-\frac{%
c_{5}^{\ast }}{c_{4}^{\ast }}=k$. Then, $c_{2}=kc_{3}$ and $c_{5}^{\ast
}=-kc_{4}^{\ast }$. From Eq. (\ref{x4}), obtain $|c_{2}|=|c_{5}|$ and $%
|c_{4}|=|c_{3}|$.

For the state $|\pi _{1}\rangle $, 3-tangle $\tau
_{123}=4|-c_{2}c_{5}+c_{3}c_{4}|^{2}$ \cite{pla-li}. When $%
c_{3}c_{4}=c_{2}c_{5}$, then $\frac{c_{2}}{c_{3}}=\frac{c_{4}}{c_{5}}$,
which contradicts that $\frac{c_{2}}{c_{3}}=-\frac{c_{5}^{\ast }}{%
c_{4}^{\ast }}$ obtained from Eq. (\ref{t1}). Therefore, $c_{3}c_{4}\neq
c_{2}c_{5}$, thus 3-tangle $\tau _{123}$ does not vanish and $|\pi
_{1}\rangle $ belongs to GHZ SLOCC class. So, under Eq. (\ref{t1}), $%
|c_{2}|=|c_{5}|,$ and $|c_{4}|=|c_{3}|$, $|\pi _{1}\rangle $ is suitable for
PSDC-3.

(C 3.1.2). $c_{2}c_{3}=0$. It cannot happen via Eqs. (\ref{t1}, \ref{x4}).

(C.3.2). $c_{4}=c_{5}=0$ \& $c_{0}c_{1}\neq 0$.

If $c_{4}=c_{5}=0$ then $c_{2}=c_{3}=0$ from Eq. (\ref{x4}). From Eq. (\ref%
{t2}), there are two subcases: $c_{6}c_{7}\neq 0$ and $c_{6}c_{7}=0$.

(C.3.2.1). $c_{6}c_{7}\neq 0$. Thus, the state is reduced to $|\pi
_{2}\rangle =c_{0}|000\rangle +c_{1}|001\rangle +c_{6}|110\rangle
+c_{7}|111\rangle $. From Eq. (\ref{t2}), obtain $\frac{c_{6}}{c_{7}}=-\frac{%
c_{1}^{\ast }}{c_{0}^{\ast }}$. Let $\frac{c_{6}}{c_{7}}=-\frac{c_{1}^{\ast }%
}{c_{0}^{\ast }}=l$. Then, $c_{6}=lc_{7}$ and $c_{1}^{\ast }=-lc_{0}^{\ast }$%
. From Eq. (\ref{x3}), obtain $|c_{0}|^{2}=|c_{7}|^{2}$ and $%
|c_{1}|^{2}=|c_{6}|^{2}$.

For the state $|\pi _{2}\rangle $, 3-tangle $\tau
_{123}=4|c_{0}c_{7}-c_{1}c_{6}|^{2}$. Assume that $c_{0}c_{7}=c_{1}c_{6}$,
then we have $\frac{c_{6}}{c_{7}}=\frac{c_{0}}{c_{1}}$, which contradicts
that $\frac{c_{6}}{c_{7}}=-\frac{c_{1}^{\ast }}{c_{0}^{\ast }}$ obtained
from Eq. (\ref{t2}). Therefore, $c_{0}c_{7}\neq c_{1}c_{6}$. Thus 3-tangle $%
\tau _{123}$ does not vanish and then $|\pi _{2}\rangle $ belongs to GHZ
class. So, $|\pi _{2}\rangle $ under that $|c_{0}|=|c_{7}|$ and $%
|c_{1}|=|c_{6}|$\ and Eq. (\ref{t2}) is suitable for PSDC-3.

(C.3.2.2). $c_{6}c_{7}=0$. It cannot happen via Eqs. (\ref{x3}, \ref{t2}).

(C.3.3.). $c_{0}=c_{5}=0$ \& $c_{1}c_{4}\neq 0$. From Eqs. (\ref{u1}, \ref%
{t2}), $c_{7}=c_{3}=0$. Then, from Eqs. (\ref{u2}, \ref{t1}), $c_{2}=c_{6}=0$%
. From Eq. (\ref{x3}), obtain $c_{1}=0$. However, $c_{1}\neq 0$. It makes a
contradiction.

(C.3.4). $c_{0}=c_{4}=0$ \& $c_{1}c_{5}\neq 0$. From Eqs. (\ref{u1}, \ref{t2}%
), $c_{7}=c_{3}=0$. Then, from Eqs. (\ref{x3}, \ref{x4}), $c_{2}c_{6}\neq 0$
and $|c_{2}|=|c_{5}|$ and $|c_{1}|=|c_{6}|$. From Eq. (\ref{a3}), obtain $%
\frac{c_{5}}{c_{6}}=-\frac{c_{2}^{\ast }}{c_{1}^{\ast }}$. Thus, obtain the
state $|\pi _{3}\rangle =c_{1}|001\rangle +c_{2}|010\rangle
+c_{5}|101\rangle +c_{6}|110\rangle $.

For $|\pi _{3}\rangle $, 3-tangle is $\tau _{123}=4|-$ $c_{2}c_{5}+$ $%
c_{1}c_{6}|^{2}$. Assume that $c_{1}c_{6}=$ $c_{2}c_{5}$. Thus, $\frac{c_{5}%
}{c_{6}}=\frac{c_{1}}{c_{2}}$, which contradicts with $\frac{c_{5}}{c_{6}}=-%
\frac{c_{2}^{\ast }}{c_{1}^{\ast }}$ obtained from Eq. (\ref{a3}).
Therefore, $c_{1}c_{6}\neq $ $c_{2}c_{5}$. Thus, 3-tangle does not vanish
and then $|\pi _{3}\rangle $ belongs to GHZ SLOCC\ class. Under Eqs. (\ref%
{a1}, \ref{a3}), and $|c_{2}|=|c_{5}|$ and $|c_{1}|=|c_{6}|$, $|\pi
_{3}\rangle $ is suitable for PSDC-3.

(C.3.5). $c_{1}=c_{5}=0$ \& $c_{0}c_{4}\neq 0$. From Eqs. (\ref{t2}, \ref{t1}%
), $c_{6}=c_{2}=0$. Then, from Eqs. (\ref{x3}, \ref{x4}), $|c_{3}|=|c_{4}|$
and $|c_{0}|=|c_{7}|$. Thus, obtain $|\pi _{4}\rangle =c_{0}|000\rangle
+c_{3}|011\rangle +c_{4}|100\rangle +c_{7}|111\rangle $. From Eq. (\ref{a3}%
), $\frac{c_{4}}{c_{7}}=-\frac{c_{3}^{\ast }}{c_{0}^{\ast }}$.

For $|\pi _{4}\rangle $, 3-tangle is $\tau
_{123}=4|c_{0}c_{7}-c_{3}c_{4}|^{2}$. Assume that $c_{0}c_{7}=c_{3}c_{4}$.
Then, $\frac{c_{4}}{c_{7}}=\frac{c_{0}}{c_{3}}$, which contradicts that $%
\frac{c_{4}}{c_{7}}=-\frac{c_{3}^{\ast }}{c_{0}^{\ast }}$ obtained from Eq. (%
\ref{a3}). Therefore, $c_{0}c_{7}\neq c_{3}c_{4}$. Thus, 3-tangle does not
vanish and then $|\pi _{4}\rangle $ belongs to GHZ SLOCC\ class. Under Eqs. (%
\ref{a1}, \ref{a3}) and $|c_{3}|=|c_{4}|$ and $|c_{0}|=|c_{7}|$, $|\pi
_{4}\rangle $ is suitable for PSDC-3.

(C.3.6). $c_{1}=c_{4}=0$ \& $c_{0}c_{5}\neq 0$. From Eqs. (\ref{t2}, \ref{t1}%
, \ref{u1}, \ref{u2}), $c_{2}=c_{3}=c_{6}=c_{7}=0$. It contradicts that $%
c_{0}c_{5}\neq 0$ via Eqs. (\ref{x3}, \ref{x4}).

(C.4). All of $c_{0},c_{1},c_{5},c_{4}$ vanish.

When $c_{0}=c_{1}=c_{5}=c_{4}=0$, from Eqs. (\ref{x3}, \ref{x4}), the case
cannot be happen.

\end{document}